\documentclass[twocolumn,amsmath,nofootinbib]{revtex4} 

\usepackage{url}
\usepackage{amssymb}
\usepackage{amsfonts}
\usepackage{amsbsy}
\usepackage{hyperref}
\usepackage{amssymb}
\usepackage{mathrsfs}
\usepackage{amsmath}
\usepackage{mathtools}
\usepackage{algorithm}
\usepackage{algorithmic}
\usepackage{tabularx}
\usepackage{enumitem}
\usepackage{mathtools}

\newcommand{\R}{\mathbb{R}}
\newcommand{\T}{\mathbfcal{T}|}

\newcommand{\E}{\mathbb{E}}
\newcommand{\h}{\hbar}
\newcommand{\p}[2]{\frac{\partial #1}{\partial #2}}

\def\ie{{\frenchspacing\it i.e.}}
\def\eg{{\frenchspacing\it e.g.}}
\def\etc{{\frenchspacing\it etc.}}
\def\rms{{\frenchspacing r.m.s.}}

\def\sign{\hbox{sign}\,}

\def\B{\textbf{B}}

\def\D{D}
\def\E{\textbf{E}}
\def\F{\textbf{F}}

\def\J{\textbf{J}}

\def\Ell{\mathcal{L}}

\def\T{\mathcal{T}}
\def\ro{{\rho}}
\def\vtheta{{\boldsymbol{\theta}}}
\def\vphi{{\boldsymbol{\phi}}}

\def\c{{\textbf{c}}}

\def\f{{\textbf{f}}}
\def\g{\textbf{g}}
\def\h{\textbf{h}}
\def\l{\ell}
\def\p{\textbf{p}}

\def\u{\textbf{u}}

\DeclareMathAlphabet\mathbfcal{OMS}{cmsy}{b}{n}

\def\x{\textbf{x}}

\def\y{\textbf{y}}

\usepackage{amsmath}
\DeclareMathOperator*{\argmax}{arg\,max}
\DeclareMathOperator*{\argmin}{arg\,min}

 \def\logplus{\log_+}
\def\DL{{\rm DL}}

\newtheorem{theorem}{Theorem}
\newtheorem{corollary}{Corollary}[theorem]

\def\spose#1{\hbox to 0pt{#1\hss}}
\def\simlt{\mathrel{\spose{\lower 3pt\hbox{$\mathchar"218$}}
   \raise 2.0pt\hbox{$\mathchar"13C$}}}
\def\simgt{\mathrel{\spose{\lower 3pt\hbox{$\mathchar"218$}}
     \raise 2.0pt\hbox{$\mathchar"13E$}}}
 \def\simpropto{\mathrel{\spose{\lower 3pt\hbox{$\mathchar"218$}}
     \raise 2.0pt\hbox{$\propto$}}}

\def\beq#1{\begin{equation}\label{#1}}
\def\eeq{\end{equation}}
\def\beqa#1{\begin{eqnarray}\label{#1}}
\def\eeqa{\end{eqnarray}}
\def\eq#1{equation~(\ref{#1})}

\def\fig#1{Figure~\ref{#1}}
\def\Fig#1{Figure~\ref{#1}}

\def\Sec#1{Section~\ref{#1}}

\parskip=\medskipamount
\begin{document}
\title{Toward an AI Physicist for Unsupervised Learning
}
\author{Tailin Wu, Max Tegmark}
\address{Dept.~of Physics \& Center for Brains, Minds \& Machines, Massachusetts Institute of Technology, Cambridge, MA 02139; tailin@mit.edu}
\begin{abstract}
We investigate opportunities and challenges for improving unsupervised machine learning using four common strategies with a long history in physics: divide-and-conquer, Occam's razor, unification and lifelong learning. Instead of using one model to learn everything, we propose a novel paradigm 
centered around the learning and manipulation of \emph{theories}, which parsimoniously predict both aspects of the future (from past observations) and the domain in which these predictions are accurate. Specifically, we propose a novel generalized-mean-loss to encourage each theory to specialize in its comparatively advantageous domain, and a differentiable description length objective to downweight bad data and ``snap" learned theories into simple symbolic formulas. Theories are stored in a ``theory hub", which continuously unifies learned theories and can propose theories when encountering new environments. We test our implementation, the toy ``AI Physicist" learning agent, on a suite of increasingly complex physics environments. From unsupervised observation of trajectories through worlds involving random combinations of gravity, electromagnetism, harmonic motion and elastic bounces, our agent typically learns faster and produces mean-squared prediction errors about a billion times smaller than a standard feedforward neural net of comparable complexity, typically recovering integer and rational theory parameters exactly. 
Our agent successfully identifies domains with different laws of motion also for a nonlinear chaotic double pendulum in a piecewise constant force field.
\end{abstract}
\date{\today}
\vspace{10mm}	

\maketitle

\section{Introduction}
\label{IntroSec}

\subsection{Motivation}

The ability to predict, analyze and parsimoniously model observations is not only central to 
physics,
but also a goal of unsupervised machine learning, which is a key frontier in artificial intelligence (AI) research 
\cite{lecun2015deep}. Despite impressive recent progress with artificial neural nets,
they still get frequently outmatched by human researchers at such modeling, suffering from two drawbacks: 
\begin{enumerate}
\item Different parts of the data are often generated by different mechanisms in different contexts.
A big model that tries to fit all the data in one environment may therefore underperform in a new environment where some mechanisms are replaced by new ones, being inflexible and inefficient at combinatorial generalization \cite{battaglia2018relational}.  
\item Big models are generally hard to interpret, and may not reveal succinct and universal knowledge such as Newton's law of gravitation that explains only some aspects of the data. The pursuit of ``intelligible intelligence" in place of inscrutable black-box neural nets is important and timely, given the growing interest in AI interpretability from AI users and policymakers, especially for AI components involved in decisions and infrastructure where trust is important \cite{russell2015research, amodei2016concrete, boden2017principles, krakovna2016increasing}.
\end{enumerate}

\def\mytab{\hglue5mm}
\begin{table}[t]
\begin{center}
\begin{tabular}{|>{\raggedright}p{2.0cm}|p{6.4cm}|}
\hline
Strategy			&Definition\\
\hline
Divide-and-	     	&Learn multiple theories each of which  \\ % Or "Break weakest link"
\mytab conquer		& specializes to fit {\it part} of the data very well\\
\hline
Occam's		&Avoid overfitting by minimizing description\\
\mytab Razor		& length, which can include 
replacing fitted constants by simple integers or fractions.\\
\hline
Unification 		&Try unifying learned theories by introducing parameters\\
\hline
Lifelong 			&Remember learned solutions and try them\\
\mytab Learning		&on future problems\\
\hline
\end{tabular}
\end{center}
\caption{AI Physicist strategies tested.
\label{table1}
}
\end{table}
To address these challenges, we will borrow from physics the core idea of a  \emph{theory}, which parsimoniously predicts both aspects of the future (from past observations) and also the domain in which these predictions are accurate. This suggests an alternative to the standard machine-learning paradigm of fitting a single big model to all the data: instead, learning small theories one by one, and gradually accumulating and organizing them. This paradigm suggests the four specific approaches summarized in Table \ref{table1}, which we combine into a toy ``AI Physicist" learning agent: To find individual theories from complex observations, we use the divide-and-conquer strategy with multiple theories and a novel generalized-mean loss that encourages each theory to specialize in its own domain by giving larger gradients for better-performing theories. To find simple theories that avoid overfitting and generalize well, we use the strategy known as Occam's razor, favoring simple theories that explain a lot, using a computationally efficient approximation of the minimum-description-length (MDL) formalism. To unify similar theories found in different environments, we use the description length for clustering and then learn a ``master theory" for each class of theories. To accelerate future learning, we use a lifelong learning strategy where learned theories are stored in a theory hub for future use.

\subsection{Goals \& relation to prior work}

The goal
of the AI Physicist learning agent presented in this paper is quite limited, and does not even remotely approach the ambition level of problem solving by human physicists. The latter is likely to be almost as challenging as artificial general intelligence, which most AI researchers guess remains decades away \cite{muller2016future,grace2018will}.
Rather, the goal of this paper is to take a very modest but useful step in that direction, combining the four physics-inspired strategies above.  

Our approach complements other work on automatic program learning, such as neural program synthesis/induction
\cite{graves2014neural,sukhbaatar2015end,reed2015neural,parisotto2016neuro,devlin2017robustfill,bramley2018learning} 
and symbolic program induction \cite{Muggleton1991,lavrac1994inductive,liang2010learning,ellis2015unsupervised,dechter2013bootstrap} and builds on prior machine-learning work on divide-and-conquer \cite{cormen2009introduction,furnkranz1999separate,ghosh2017divide}, network simplification \cite{rissanen1978modeling,hassibi1993second,suzuki2001simple,grunwald2005advances,han2015deep,han2015learning} and continuous learning \cite{kirkpatrick2017overcoming,li2017learning,lopez2017gradient,nguyen2017variational}. It is often said that babies are born scientists, and there is arguably evidence for use of all of these four strategies during childhood development as well \cite{bramley2018learning}.

There has been significant recent progress on AI-approaches specifically linked to physics, including
physical scene understanding \cite{yildirim2018neurocomputational},
latent physical properties \cite{zheng2018unsupervised,battaglia2016interaction,chang2016compositional},
learning physics simulators \cite{watters2017visual},
physical concept discovery \cite{iten2018discovering},
an intuitive physics engine  \cite{lake2017building},
and
the ``Sir Isaac" automated adaptive inference agent \cite{daniels2015automated}.
Our AI Physicist is different and complementary in two fundamental ways that loosely correspond to the two motivations on the first page: 
\begin{enumerate}
\item All of these papers learn one big model to fit all the data. In contrast, the AI Physicist learns many small models applicable in different domains, using the divide-and-conquer strategy. 
\item Our primary focus is not 
on making approximate predictions or discovering latent variables, but on 
near-exact predictions and complete intelligibility. From the former perspective, it is typically irrelevant if a model parameter changes by a tiny amount, but from a physics perspective, one is quite interested to learn that gravity weakens like distance to the power $2$ rather than $1.99999314$.
\end{enumerate}

We share this focus on intelligibility with a long tradition of research on computational scientific discovery \citep{dvzeroski2007computational}, including the Bacon system \cite{langley1981data} and its successors \cite{langley1989data}, which induced physical laws from observations and which also used a divide-and-conquer strategy. Other work has extended this paradigm to support discovery of differential equation models from multivariate time series \cite{dzeroski1995discovering,bradley2001reasoning,langley2003robust,langley2015heuristic}.

The rest of this paper is organized as follows. In Section \ref{sec:method}, we introduce the architecture of our AI Physicist learning agent, and the algorithms implementing the four strategies. We present the results of our numerical experiments using a suite of physics environment benchmarks in Section \ref{sec:numerical_experiments}, and discuss our conclusions in Section IV, delegating supplementary technical details to a series of appendices.

\section{Methods}
\label{sec:method}

Unsupervised learning of regularities in time series can be viewed as a supervised learning problem of predicting the future from the past.
This paper focuses on the task of predicting the next state vector $\y_t\in \R^d$ in a sequence from the
concatenation $\x_t =(\y_{t-T},...,\y_{t-1})$  of the last $T$ vectors. 
 However, our AI Physicist formalism applies more generally to learning any function $\R^M\mapsto\R^N$ from examples.
In the following we first define \emph{theory}, then introduce a unified AI Physicist architecture implementing the four aforementioned strategies.

\subsection{Definition of Theory}

A theory $\mathcal{T}$ is a 2-tuple $(\f, c)$, where $\f$ is a prediction function that predicts $\y_t$ when $\x_t$ is within the theory's domain, and $c$ is a domain sub-classifier which takes $\x_t$  as input and outputs a logit of whether $\x_t$ is inside this domain. When multiple theories are present, the sub-classifier $c$'s outputs are concatenated and fed into a softmax function, producing probabilities for which theory is 
applicable. 
Both $\f$ and $c$ can be implemented by a neural net or symbolic formula,
and can be set to learnable during training and fixed during prediction/validation.

This definition draws inspirations from physics theories (conditional statements), such as ``a ball not touching anything (\emph{condition}) with vertical velocity and height $(v_0,h_0)$ will a time $t$ later have $\y\equiv(v,h)=(v_0-gt,h_0+v_0 t -gt^2/2)$ (\emph{prediction function})".
For our AI Physicist, theories constitute its ``atoms" of learning, as well as the building blocks for higher-level manipulations.

\subsection{AI Physicist Architecture Overview}

\begin{figure}[pbt]
\centerline{\includegraphics[width=80mm]{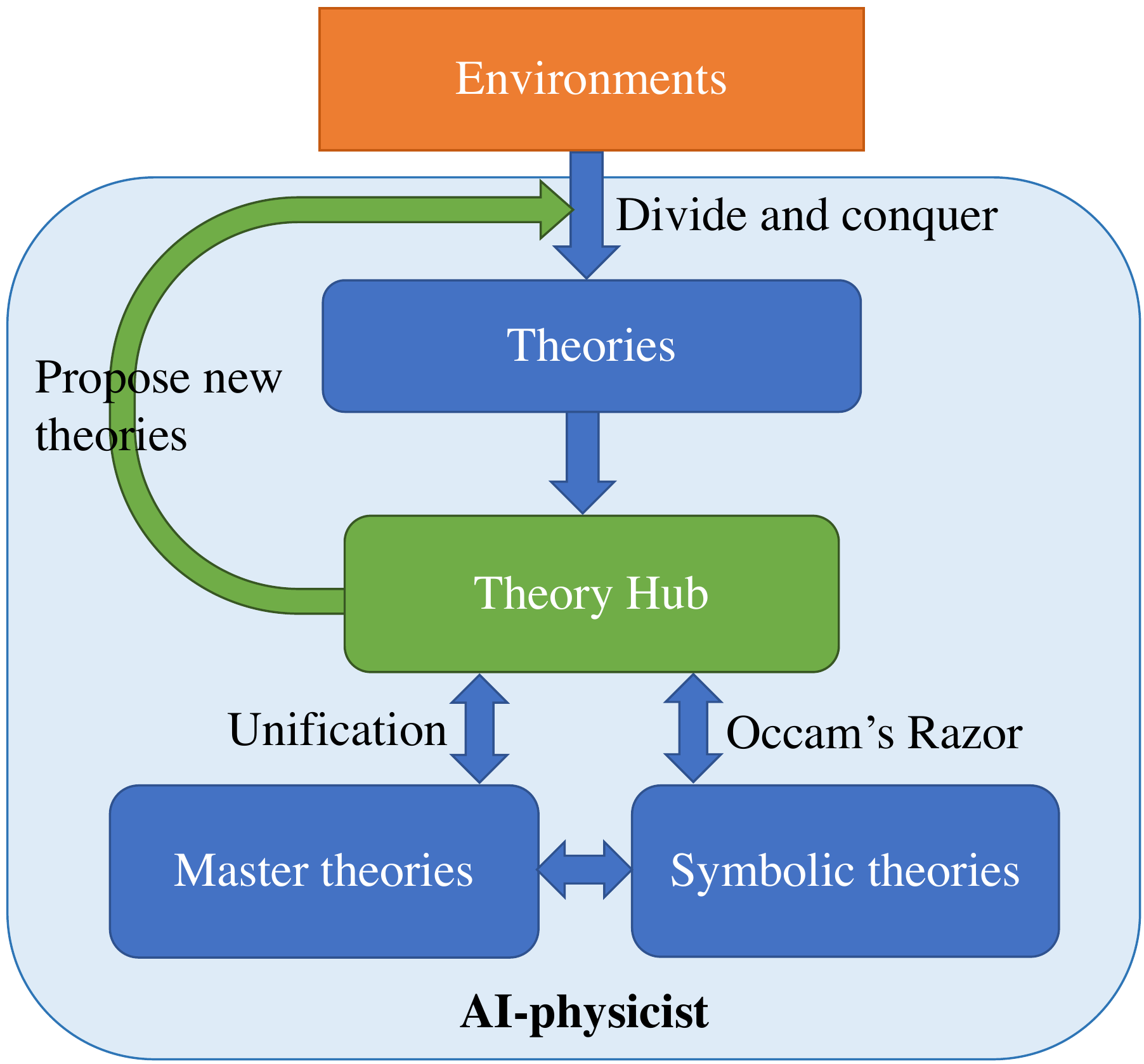}}
\caption{
AI Physicist Architecture
\label{fig:AI_physicist_architecture}
}
\end{figure}

Figure \ref{fig:AI_physicist_architecture} illustrates the architecture of the AI Physicist learning agent. At the center is a theory hub which stores the learned and organized theories. When encountering a new environment, the agent first inspects the hub and proposes old theories that help account for parts of the data as well as randomly initialized new theories for the rest of the data. All these theories are trained via our divide-and-conquer strategy, first jointly with our generalized-mean loss then separately to fine-tune each theory in its domain (section \ref{sec:divide_and_conquer}). Successful theories along with the corresponding data are added to the theory hub.

The theory hub has two organizing strategies: (1) Applying Occam's razor, it snaps the learned theories, in the form of neural nets, into simpler symbolic formulas (section \ref{sec:Occams_Razor}). (2) Applying unification, it clusters and unifies the symbolic theories into master theories (section \ref{sec:unification}). The symbolic and master theories can be added back into the theory hub, improving theory proposals for new environments. The detailed AI Physicist algorithm is presented in a series of appendices. It has polynomial time complexity, as detailed in Appendix~\ref{ComplexitySec}.

\subsection{Divide-and-Conquer}
\label{sec:divide_and_conquer}

Conventionally, a function $\f$ mapping $\x_t\mapsto\y_t$  is learned by parameterizing
$\f$ by some parameter vector $\vtheta$ that is adjusted to minimize a loss (empirical risk)
\beq{expected_risk}
\Ell\equiv\sum_t \ell[\f(\x_t), \y_t],
\eeq
where $\ell$ is some non-negative distance function quantifying how far each prediction is from the target, typically satisfying $\ell(\y,\y)=0$.
In contrast, a physicist observing an unfamiliar environment  does typically {\it not} try to predict everything with one model, instead starting with an easier question: is there any part or aspect of the world that can be described? 
For example, when Galileo famously tried to model the motion of swinging lamps in the Pisa cathedral, he completely ignored everything else, and made no attempts to simultaneously predict the behavior of sound waves, light rays, weather, or subatomic particles.
In this spirit, we allow multiple competing theories $\mathbfcal{T}=\{\T_i\}=\{(\f_i, c_i)\}$, $i=1,2,...M$, to specialize in different domains, with a novel generalized-mean loss
\beq{generalized_mean_risk_empirical}
\Ell_\gamma\equiv\sum_t\left(\frac{1}{M}\sum_{i=1}^M \ell[\f_i(\x_t), \y_t]^\gamma\right)^{1/\gamma}
\eeq

When $\gamma<0$, the loss $\Ell_\gamma$
will be dominated by whichever prediction function $\f_i$ fits each data point best. This dominance is controlled by $\gamma$, with 
$\Ell_\gamma\to \min_i \ell[\f_i(\x_t),\y_t]$ in the limit where $\gamma\to-\infty$.
This means that the best way to minimize $\Ell_\gamma$ is for each $\f_i$ to specialize by further improving its accuracy for the data points where it already outperforms the other theories.
The following Theorem \ref{thm:theorem_gradient} formalizes the above intuition, stating that under mild conditions for the loss function $\ell(\cdot,\cdot)$, the generalized-mean loss gives larger gradient w.r.t.~the error $|\hat{\y}_t-\y_t|$ for theories that perform better, so that a gradient-descent loss minimization encourages specialization. 

\begin{theorem}
\label{thm:theorem_gradient}
Let $\hat{\y}^{(i)}_t\equiv \f_i(\x_t)$ denote the prediction of the target $\y_t$ by the function $\f_i$, $i=1,2,...M$.
Suppose that $\gamma<0$ and $\ell(\hat{\y}_t, \y_t) = \ell(|\hat{\y}_t - \y_t|)$ for a monotonically increasing function 
$\ell(u)$ that vanishes on $[0,u_0]$ for some $u_0\geq 0$, with $\ell(u)^\gamma$ differentiable and strictly convex for $u>u_0$. \\
Then if $0<\ell(\hat{\y}_t^{(i)}, \y_t) < \ell(\hat{\y}_t^{(j)}, \y_t)$, we have 
\beq{gradient_greater}
\left|\frac{\partial\Ell_\gamma}{\partial u^{(i)}_t}\right| > \left|\frac{\partial \Ell_\gamma}{\partial u^{(j)}_t}\right|,
\eeq
 where $u^{(i)}_t\equiv|\hat{\y}_t^{(i)} - \y_t|$.
\end{theorem}
Appendix \ref{proof_theorem_gradient} gives the proof, and also shows
that this theorem applies to mean-squared-error (MSE) loss $\ell(u)=u^2$,
mean-absolute-error loss $\ell(u)=|u|$, Huber loss and our description-length loss from the next section.

We find empirically that the simple choice $\gamma=-1$ works quite well, striking a good balance between encouraging specialization for the best theory and also giving some gradient for theories that currently perform slightly worse. We term  this choice $\Ell_{-1}$ the ``harmonic loss", because it corresponds to the harmonic mean of the losses for the different theories. Based on the harmonic loss, we propose an unsupervised differentiable divide-and-conquer (DDAC) algorithm (Alg.~\ref{alg:divide_and_conquer} in Appendix~\ref{DivideAndConquerAlgo}) that simultaneously learns prediction functions $\{\f_i\}$ and corresponding domain classifiers $\{c_i\}$ from observations.

Our DDAC method's combination of multiple prediction modules into a single prediction is reminiscent of AdaBoost \cite{freund1997decision}.
While AdaBoost gradually upweights those modules that best predict {\it all} the data, DDAC instead 
identifies complementary modules that each predict some {\it part} of the data best, and encourages these modules to simplify and improve by specializing on these respective parts.

\subsection{Occam's Razor}
\label{sec:Occams_Razor}

The principle of  Occam's razor, that simpler explanations are better, is quite popular among physicists.
This preference for parsimony helped dispense with phlogiston, aether and other superfluous concepts.

Our method therefore incorporates the minimum-description-length (MDL) formalism \cite{rissanen1978modeling,grunwald2005advances}, which provides an elegant mathematical implementation of Occam's razor. It is rooted in Solomonoff's theory of inference \cite{solomonoff1964formal} and is linked to Hutter's AIXI approach to artificial general intelligence \cite{hutter2000theory}. 
The description length (DL) of a dataset $\D$ is defined as the number of bits required to describe it.
For example, if regularities are discovered that enable data compression, then the corresponding description length is defined as 
the number of bits of the program that produces $\D$ as its output (including both the code bits and the compressed data bits).
In our context of predicting a time series, this means that the description length is the number of bits required to describe the theories used plus the number of bits required to store all prediction errors.
Finding the optimal data compression and hence computing the MDL is a famous hard problem that involves searching an exponentially large space, but any discovery reducing the description length is a step in the right direction, and provably avoids the overfitting problem that plagues many alternative machine-learning strategies \cite{rissanen1978modeling,grunwald2005advances}.

The end-goal of the AI Physicist is to discover theories $\mathbfcal{T}$ minimizing the total description length, given by
\beq{description_length}
\DL(\mathbfcal{T},\D)=\DL(\mathbfcal{T})+\sum_t\DL(\u_t),
\eeq
where $\u_t=\hat{\y}_t-\y_t$ is the prediction error at time step $t$. By discovering simple theories that can each account for parts of the data very well, the AI Physicist strives to make both $\DL(\mathbfcal{T})$ and $\sum_t\DL(\u_t)$ small.

Physics has enjoyed great success in its pursuit of simpler theories using rather vague definitions of simplicity. In the this spirit, we choose to compute the description length DL not exactly, but using an approximate heuristic that is numerically efficient, and significantly simpler than more precise versions such as \cite{rissanen1983universal},
paying special attention to rational numbers since they are appear in many physics theories.
We compute the DL of both theories $\T$ and prediction errors $\u_t$ as the sum of the DL of all numbers that specify them, using the following conventions for the DL of integers, rational numbers and real numbers. Our MDL implementation differs from popular machine-learning approaches whose goal is efficiency and generalizability  \cite{Hinton:1993:KNN:168304.168306,han2015deep,blierdescription} rather than intelligibility.

\begin{figure}[pbt]
\centerline{\includegraphics[width=88mm]{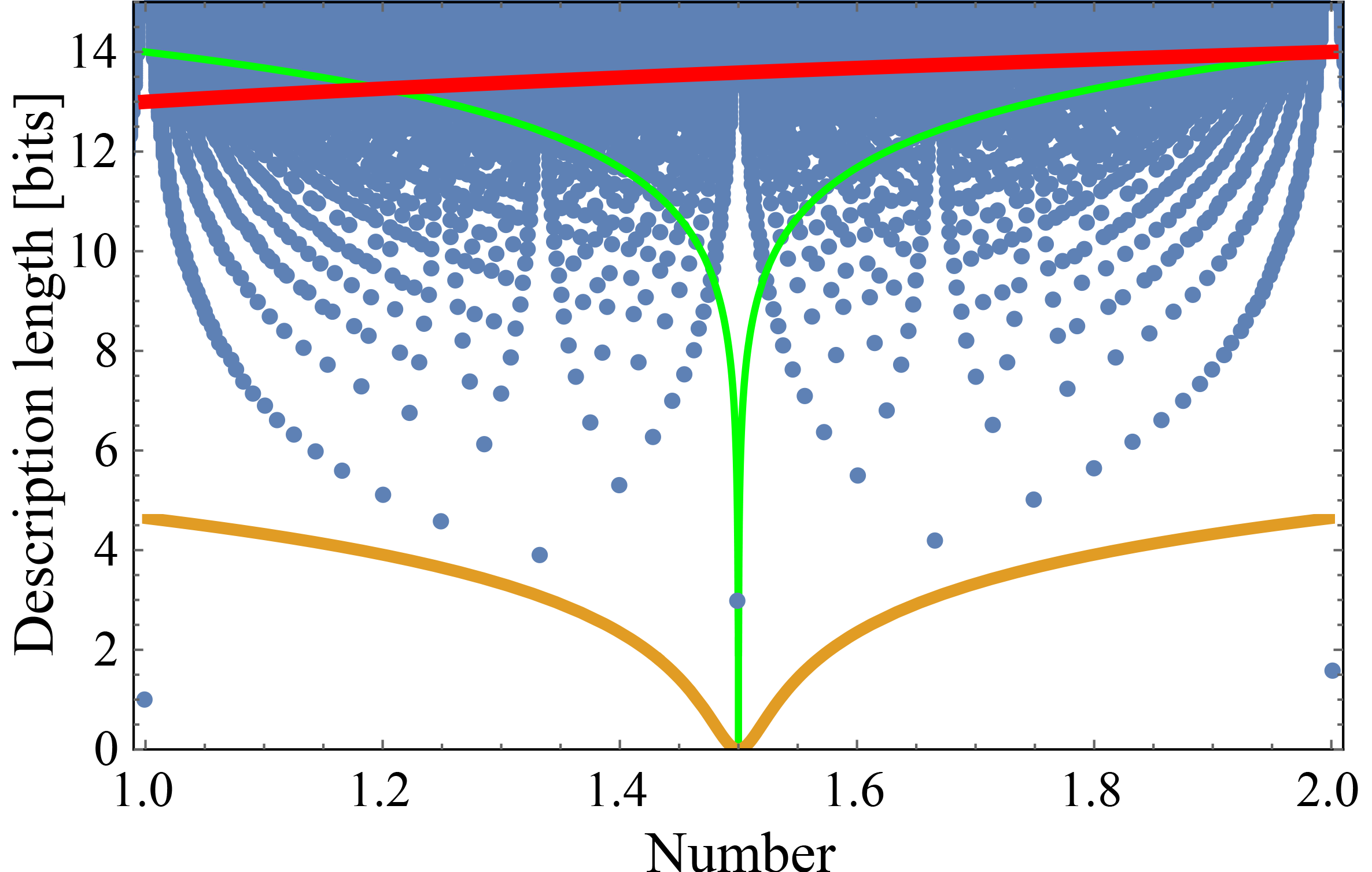}}
\caption{The description length $\DL$ is shown for real numbers $p$ with $\epsilon=2^{-14}$ (rising curve) and for 
rational numbers (dots).
Occam's Razor favors lower DL, and our MDL rational approximation of a real parameter $p$ is the lowest point after taking these ``model bits" specifying the approximate parameter and adding the ``data bits" $\Ell$ required to specify the prediction error made.
The two symmetric curves illustrate the simple example where $\Ell=\logplus\left({x-x_0\over\epsilon}\right)$
for $x_0=1.4995$, $\epsilon=2^{-14}$ and $0.02$, respectively.}
\label{RationalComplexityFig}
\end{figure}
The number of binary digits required to specify a natural number $n=1,2,3,...$ is approximately $\log_2 n$, so we define 
$\DL(n)\equiv\log_2 n$ for natural numbers. 
For an integer $m$, we define
\beq{IntegerDLeq}
\DL(m)\equiv %\DL(1+|m|)= 
\log_2 (1+|m|).
\eeq
For a rational number $q=m/n$, the description length is the sum of that for its integer numerator and (natural number) denominator, as illustrated in \fig{RationalComplexityFig}:
\beq{RationalDLeq}
\DL\left({m\over n}\right)=\log_2[(1+|m|)n].
\eeq
For a real number $r$ and a numerical precision floor $\epsilon$, we define
\beq{RealDLeq}
\DL(r)=\logplus\left({r\over\epsilon}\right),
\eeq
where the function
\beq{LogplusDefEq}
\logplus(x)\equiv{1\over 2}\log_2\left(1+x^2\right)
\eeq
is plotted in \fig{RationalComplexityFig}.
Since  $\logplus(x)\approx\log_2 x$ for $x\gg1$, 
$\DL(r)$ is approximately the description length of  the integer closest to $r/\epsilon$.
Since  $\logplus(x)\simpropto x^2$ for $x\ll 1$, $\DL(r)$ simplifies to a quadratic (mean-squared-error) loss function below the numerical precision, which will prove useful 
below.\footnote{Natural alternative definitions of $\logplus(x)$ include $\log_2\left(1+|x|\right)$, $\log_2\max(1,|x|)$, $(\ln 2)^{-1}\sinh^{-1}|x|$
and $(2\ln 2)^{-1}\sinh^{-1}(x^2)$. Unless otherwise specified, we choose $\epsilon=2^{-32}$ in our experiments.}

Note that as long as all prediction absolute errors $|u_i|\gg\epsilon$ %($u_i$ are elements of an error vector $\u$) 
for some dataset, 
minimizing the total description length $\sum_i \DL(u_i)$ instead of the MSE $\sum_i u_i^2$
corresponds to minimizing the geometric mean instead of the arithmetic mean of the squared errors, which encourages focusing more on improving already well-fit points. 
$\sum_i \DL(u_i)$ drops by 1 bit whenever one prediction error is halved, which is can typically be achieved by fine-tuning the fit for many valid data points that are already well predicted while increasing DL for bad or extraneous points at most marginally.

\begin{figure}[pbt]
\centerline{\includegraphics[width=88mm]{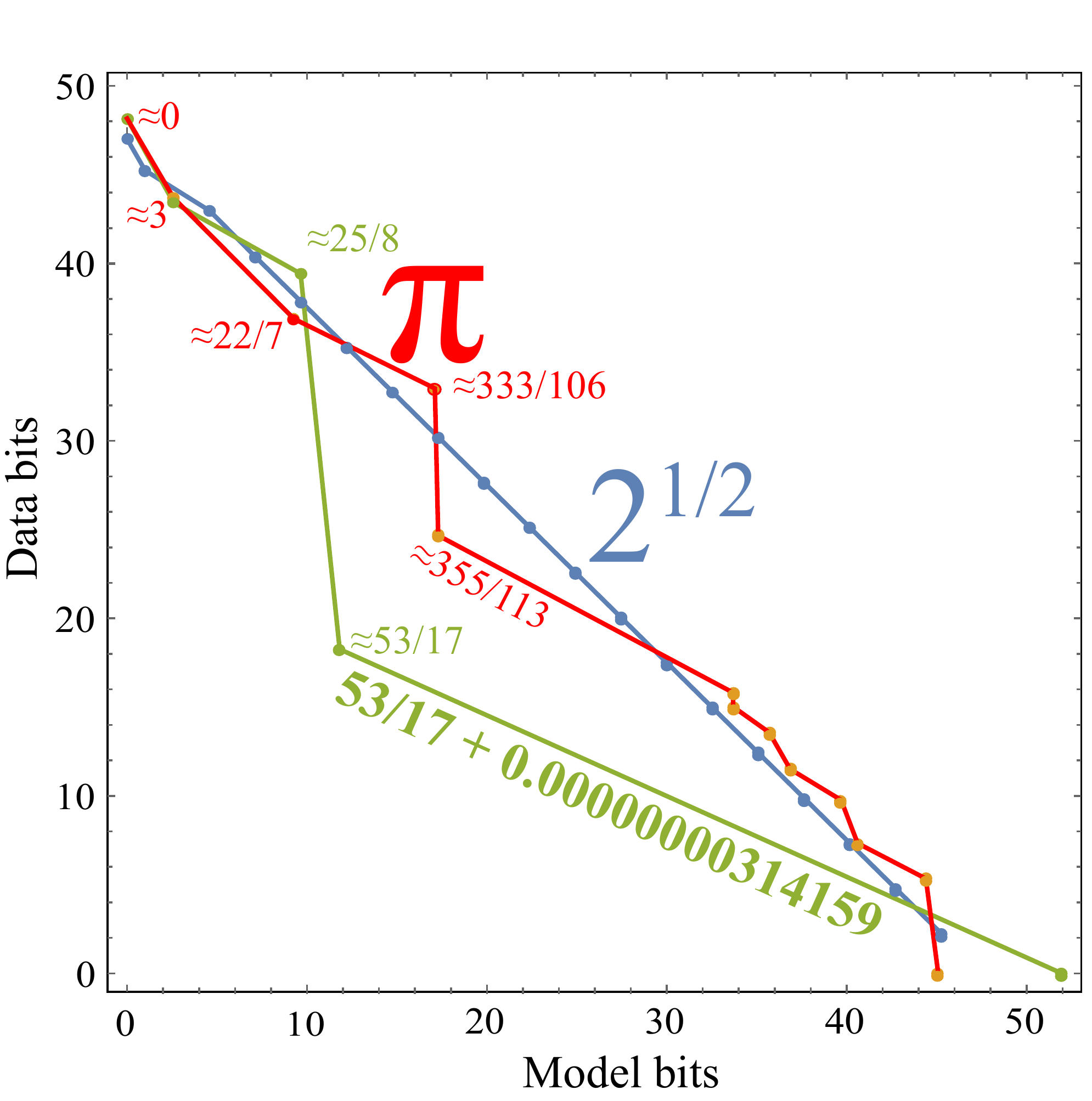}}
\caption{Illustration of our minimum-description-length (MDL) analysis 
of the parameter vector $\p=\{\pi,\sqrt{2},3.43180632382353\}$.
We approximate each real number $r$ as a fraction $a_k/b_k$ using the first $k$ terms of its  continued fraction expansion, and for each integer $k=1,...$, we plot the number of ``data bits" required to encode the prediction error $r-a_k/b_k$ to 14 decimal places versus the number of ``model bits" required to encode the rational approximation $a_k/b_k$, as described in the text. We then select the point with smallest bit sum (furthest down/left from the diagonal) as our first approximation candidate to test. 
Generic irrational numbers are incompressible; the total description length (model bits+data bits) is roughly independent of $k$ as is seen for $\pi$ and $\sqrt{2}$, corresponding to a line of slope $-1$ around which there are small random fluctuations. In contrast, the green/light grey curve (bottom) is for a parameter that is anomalously close to a rational number, and the curve reveals this by the approximation $53/17$ reducing the total description length (model$+$data bits) by about 16 bits.
\label{NumberMDLfig}
}
\end{figure}

For numerical efficiency, our AI Physicist minimizes the description length of 
\eq{description_length} in two steps:
1) All model parameters are set to trainable real numbers, and the DDAC algorithm is applied to minimize the harmonic loss $\Ell_{-1}$ with 
$\ell(\u)\equiv\sum_i \DL(u_i)$ using \eq{RealDLeq} and the annealing procedure for the precision floor described in Appendix~\ref{DivideAndConquerAlgo}.
2) Some model parameters are replaced by rational numbers as described below, followed by re-optimization of the other parameters. 
The idea behind the second step is that if a physics experiment or neural net training produces a parameter $p=1.4999917$, it would be natural to interpret this as a hint, and to check if $p=3/2$ gives an equally acceptable fit to the data, reducing total DL.
We implement step 2 using continued fraction expansion as described in Appendix \ref{appendix:OccamsRazorAlgo} and illustrated in \fig{NumberMDLfig}.

\subsection{Unification}
\label{sec:unification}

Physicists aspire not only to find simple theories that explain aspects of the world accurately, but also to discover underlying similarities between theories and {\it unify} them. 
For example, when James Clerk Maxwell corrected and unified four key formulas describing electricity and magnetism into his eponymous equations ($\mathrm{d}\F=0$, 
$\mathrm{d}\star \F=\J$ in differential form notation), he revealed the nature of light and enabled the era of wireless communication.

Here we make a humble attempt to automate part of this process. The goal of the unification is to output a master theory $\mathscr{T}=\{(\f_\p,\cdot)\}$,  such that varying the parameter vector $\p\in\R^n$ can generate a continuum of theories $(\f_\p,\cdot)$ including previously discovered ones.
For example, Newton's law of gravitation can be viewed as a master theory unifying the gravitational force formulas around different planets by introducing a parameter $p$ corresponding to planet mass. Einstein's special relativity can be viewed as a master theory unifying the approximate formulas for $v\ll c$ and $v\approx c$ motion. 

We perform unification by first computing the description length $\text{dl}^{(i)}$ of the prediction function $\f_i$ (in  symbolic form) for each theory $i$ and performing clustering on $\{\text{dl}^{(i)}\}$. Unification is then achieved by discovering similarities and variations between the symbolic formulas in each cluster, retaining the similar patterns, and introducing parameters in place of the parameters that vary as detailed in Appendix \ref{UnificationAlgo}.

\subsection{Lifelong Learning}

Isaac Newton once said ``If I have seen further it is by standing on the shoulders of giants", emphasizing the utility of  building on past discoveries. At a more basic level, our past experiences enable us humans to model new environments much faster than if we had to re-acquire all our knowledge from scratch. We therefore embed a 
lifelong learning strategy
into the architecture of the AI Physicist. As shown in Fig. \ref{fig:AI_physicist_architecture} and Alg.~\ref{alg:overall_algorithm}, the theory hub stores successfully learned theories, organizes them with our Occam's razor and unification algorithms (reminiscent of what humans do while dreaming and reflecting), and when encountering new environments, uses its accumulated knowledge to propose new theories that can explain parts of the data. This both
ensures that past experiences are not forgotten and enables faster learning in novel environments. The detailed algorithms for proposing and adding theories are in Appendix \ref{appendix:theory_proposal_adding}.

\section{Results of  Numerical Experiments}
\label{sec:numerical_experiments}

\subsection{Physics Environments}

\begin{figure}[phbt]
\centerline{\includegraphics[width=86mm]{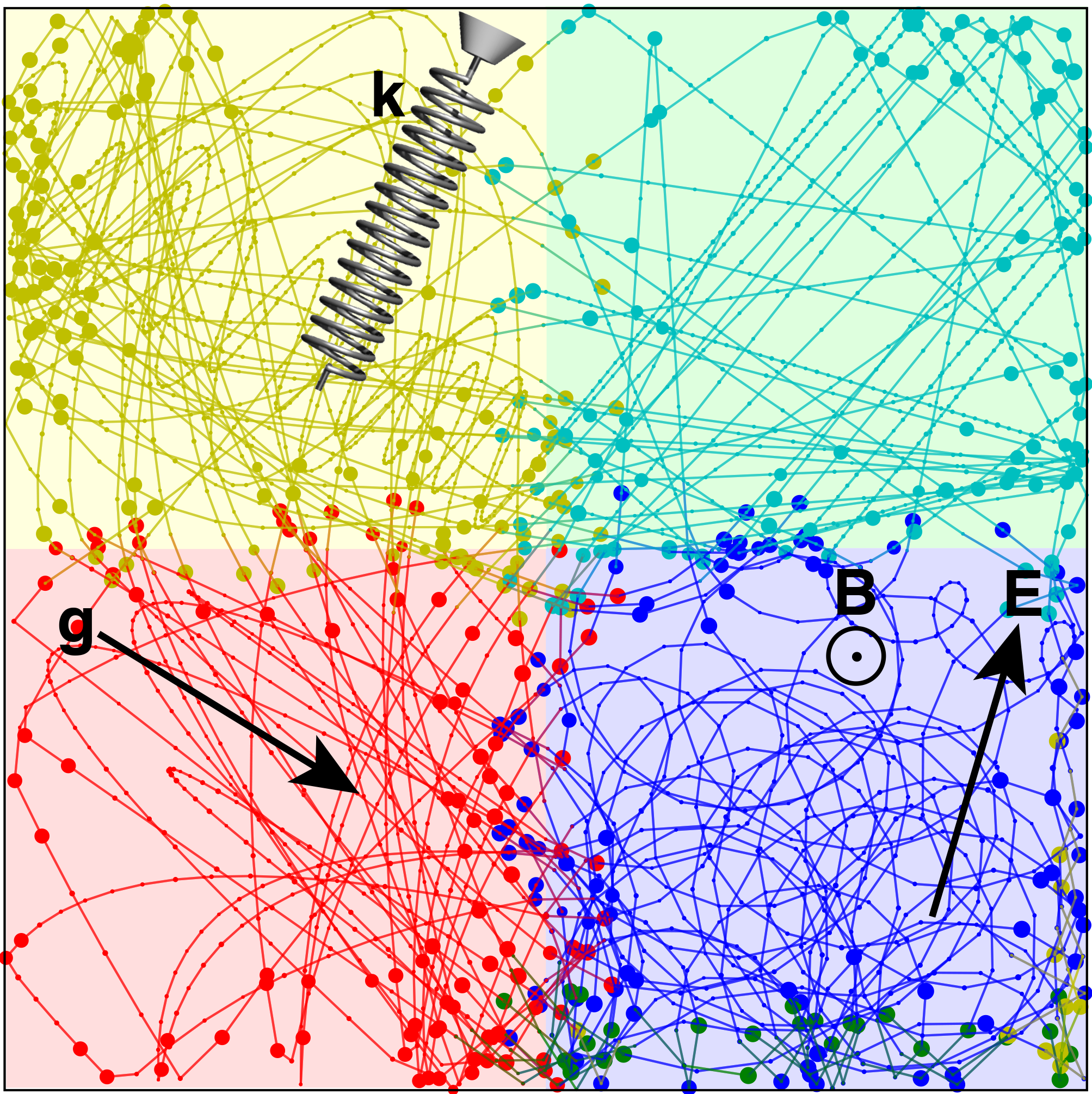}}
\caption{In this sample mystery world, a ball moves through a harmonic potential (upper left quadrant), a gravitational field (lower left) and an electromagnetic field (lower right quadrant) and bounces elastically from four walls. 
The only input to the AI Physicist is the sequence of dots (ball positions); the challenge is to learn all boundaries and laws of motion (predicting each position from the previous two). 
The color of each dot represents the domain into which it is classified by 
$\c$,
and its area represents the description length of the error with which its position is predicted ($\epsilon=10^{-6}$) after the DDAC
(differentiable divide-and-conquer) algorithm; the AI Physicist tries to minimize the total area of all dots.
\label{WorldExampleFig}
}
\end{figure}

We test our algorithms on two suites of benchmarks, each with increasing complexity. In all cases, the goal is to predict the two-dimensional motion as accurately as possible. One suite involves chaotic and highly nonlinear motion of a charged double pendulum in two adjacent electric fields. The other suite involves balls affected by gravity, electromagnetic fields, springs and bounce-boundaries, as exemplified in \fig{WorldExampleFig}.
Within each spatial region, the force corresponds to a potential energy function $V\propto (ax+by+c)^n$ for some constants $a$, $b$, $c$, where 
$n=0$ (no force), 
$n=1$ (uniform electric or gravitational field), 
$n=2$ (spring obeying Hooke's law)
or $n=\infty$ (ideal elastic bounce), and optionally involves also a uniform magnetic field.
The environments are summarized in Table~\ref{DetailedResultsTable2}.

\subsection{Numerical Results}

In the mystery world example of \fig{WorldExampleFig}, after the DDAC algorithm~\ref{alg:divide_and_conquer} taking the sequence of coordinates as the only input, we see that the AI Physicist has learned to simultaneously predict the future position of the ball from the previous two, and classify 
without external supervision the observed inputs into four big physics domains. The predictions are seen to be more accurate deep inside the domains (tiny dots) than near boundaries (larger dots) where transitions and bounces create small domains with laws of motion that are harder to infer because of complexity and limited data. Because these small domains can be automatically inferred and eliminated once the large ones are known as described in Appendix~\ref{DomainElimination}, all accuracy benchmarks quoted below refer to points in the large domains only.

After DDAC, the AI Physicist performs Occam's-razor-with-MDL (Alg.~\ref{alg:MDL_simplification}) 
on the learned theories. As an example, it discovers that the motion deep inside the lower-left quadrant obeys the difference equation parameterized by a learned 3-layer neural net, which after the first collapseLayer transformation simplifies to
\beqa{RawDifferenceEq}
\hat{\y}_t&=&\left(
\begin{tabular}{rrrr}
-0.99999994&0.00000006&1.99999990&-0.00000012\\ 
-0.00000004&-1.0000000&0.00000004&2.00000000 
\end{tabular}
\right)\x_t\nonumber\\
&+& 
\left(
\begin{tabular}{r}
 0.01088213\\ 
-0.00776199
\end{tabular}
\right),
\eeqa

with $\DL(\f)=212.7$ and $\sum_t\DL(\u_t)=2524.1$.
The snapping stage thereafter simplifies this to
\beq{RawDifferenceEqSnapped}
\hat{\y}_t=\left(
\begin{tabular}{rrrr}
-1&0&2&0\\ 
0&-1&0&2 
\end{tabular}
\right)\x_t
+ 
\left(
\begin{tabular}{r}
 0.010882\\ 
-0.007762 
\end{tabular}
\right).
\eeq
which has lower description length in both model bits ($\DL(\f)=55.6$) and data bits ($\sum_t\DL(\u_t)=2519.6$) and gets transformed to the symbolic expressions
\begin{eqnarray}
\hat{x}_{t+2}&=&2 x_{t+1}-x_t+0.010882,\nonumber\\
\hat{y}_{t+2}&=&2 y_{t+1}-y_t-0.007762,
\end{eqnarray}
where we have writen the 2D position vector $\y=(x,y)$ for brevity.
During unification (Alg.~\ref{UnificationAlgo}), the AI Physicist discovers multiple clusters of theories based on the DL of each theory, where one cluster has DL ranging between 48.86 and 55.63, which it unifies into a master theory $\f_\p$ with
\beqa{GravityUnifiedEq}
\hat{x}_{t+2}&=&2 x_{t+1}-x_t+p_1,\nonumber\\
\hat{y}_{t+2}&=&2 y_{t+1}-y_t+p_2,
\eeqa
effectively discovering a ``gravity" master theory out of the different types of environments it encounters.
If so desired, the difference equations~(\ref{GravityUnifiedEq}) can be automatically generalized to the more familiar-looking differential equations 
\begin{eqnarray}
\ddot x&=&g_x,\nonumber\\
\ddot y&=&g_y,\nonumber
\end{eqnarray}
where $g_i\equiv p_i(\Delta t)^2$,
and both the Harmonic Oscillator Equation and Lorentz Force Law of electromagnetism can be analogously auto-inferred from other master theories learned.

Many mystery domains in our test suite involve laws of motion whose parameters include both rational and irrational numbers. To count a domain as ``solved" below, we use the very stringent requirement that any rational numbers (including integers) must be discovered {\it exactly}, while irrational numbers must be recovered with accuracy $10^{-4}$.

We apply our AI Physicist to 40 mystery worlds in sequence (Appendix~\ref{DetailedResultsSec}). After this training, we apply it to a suite of 40 additional worlds to test how it learns different numbers of examples.
The results 
are shown in tables~\ref{DetailedResultsTable}
and~\ref{DetailedResultsTable2}, and  
Table~\ref{ResultsSummaryTable} summarizes these results using the median over worlds.
For comparison, we also show results for two simpler agents with similar parameter count: a ``baseline" agent consisting of a three-layer feedforward MSE-minimizing leakyReLU network and a ``newborn" AI Physicist that has not seen any past examples and therefore cannot benefit from the lifelong learning strategy.

We see that the newborn agent outperforms baseline on all the tabulated measures, and that the AI Physicist does still better. 
Using all data, the Newborn agent and AI Physicist are able to predict
with mean-squared prediction error below $10^{-13}$, more than nine orders of magnitude below baseline. Moreover, the Newborn and AI Physicist agents are able to simultaneously learn the domain classifiers with essentially perfect accuracy, without external supervision. Both agents are able to solve above 90\% of all the 40 mystery worlds according to our stringent criteria.

The main advantage of the AI Physicist over the Newborn agent is seen to be its learning speed, attaining given accuracy levels faster, especially during the early stage of learning. Remarkably, for the subsequent 40 worlds, the AI Physicist reaches 0.01 MSE within 35 epochs using as little as 1\% of the data, performing almost as well as with 50\% of the data much better than the Newborn agent. This illustrates that the lifelong learning strategy enables the AI Physicist to learn much faster in novel environments with less data.
This is much like an experienced scientist can solve new problems way faster than a beginner by building on prior knowledge about similar problems.
  
\begin{table}
\begin{center}
\begin{tabular}{|l|r|r|r|}
\hline
Benchmark			&Baseline		&Newborn		&AI Physicist\\
\hline
$\log_{10}$ mean-squared error 	&-3.89	&-13.95		&-13.88\\
Classification accuracy	&67.56\%		&100.00\%	&100.00\%\\
Fraction of worlds solved	&0.00\%		&90.00\%		&92.50\%\\
Description length for $\f$		&11,338.7		&198.9 		&198.9\\
Epochs until $10^{-2}$ MSE		&95			&83 			&15\\
Epochs until $10^{-4}$ MSE	&6925		&330		&45\\
Epochs until $10^{-6}$ MSE&$\infty$		&5403		&3895\\
Epochs until $10^{-8}$ MSE&$\infty$		&6590		&5100\\
\hline
$\log_{10}$ MSE		&			&			&\\
$\>\>$ using 100\% of data 	&-3.78		&-13.89		&-13.89\\
$\>\>$ using 50\% of data 	&-3.84		&-13.76		&-13.81	\\
$\>\>$ using 10\% of data 	&-3.16		&-7.38		&-10.54\\
$\>\>$ using 5\% of data 	&-3.06		&-6.06 		&-6.20\\
$\>\>$ using 1\% of data 	&-2.46		&-3.69 		&-3.95\\
\hline
Epochs until $10^{-2}$ MSE 	&			& 			&	\\
$\>\>$ using 100\% of data 	&95		&80	&15\\
$\>\>$ using 50\% of data 	&190		&152.5	&30\\
$\>\>$ using 10\% of data 	&195		&162.5	&30\\
$\>\>$ using 5\% of data 	&205		&165	&30\\
$\>\>$ using 1\% of data 	&397.5		&235	&35\\
\hline
\end{tabular}
\end{center}
\caption{Summary of numerical results, taking the median over 40 mystery environments
from Table~\ref{DetailedResultsTable} (top part) and on 40 novel environments with varying fraction of random examples (bottom parts), where each world is run with 10 random initializations and taking the best performance.
Accuracies refer to big regions only.
\label{ResultsSummaryTable}
}
\end{table}

\begin{figure}[phbt]
\centerline{\includegraphics[width=86mm]{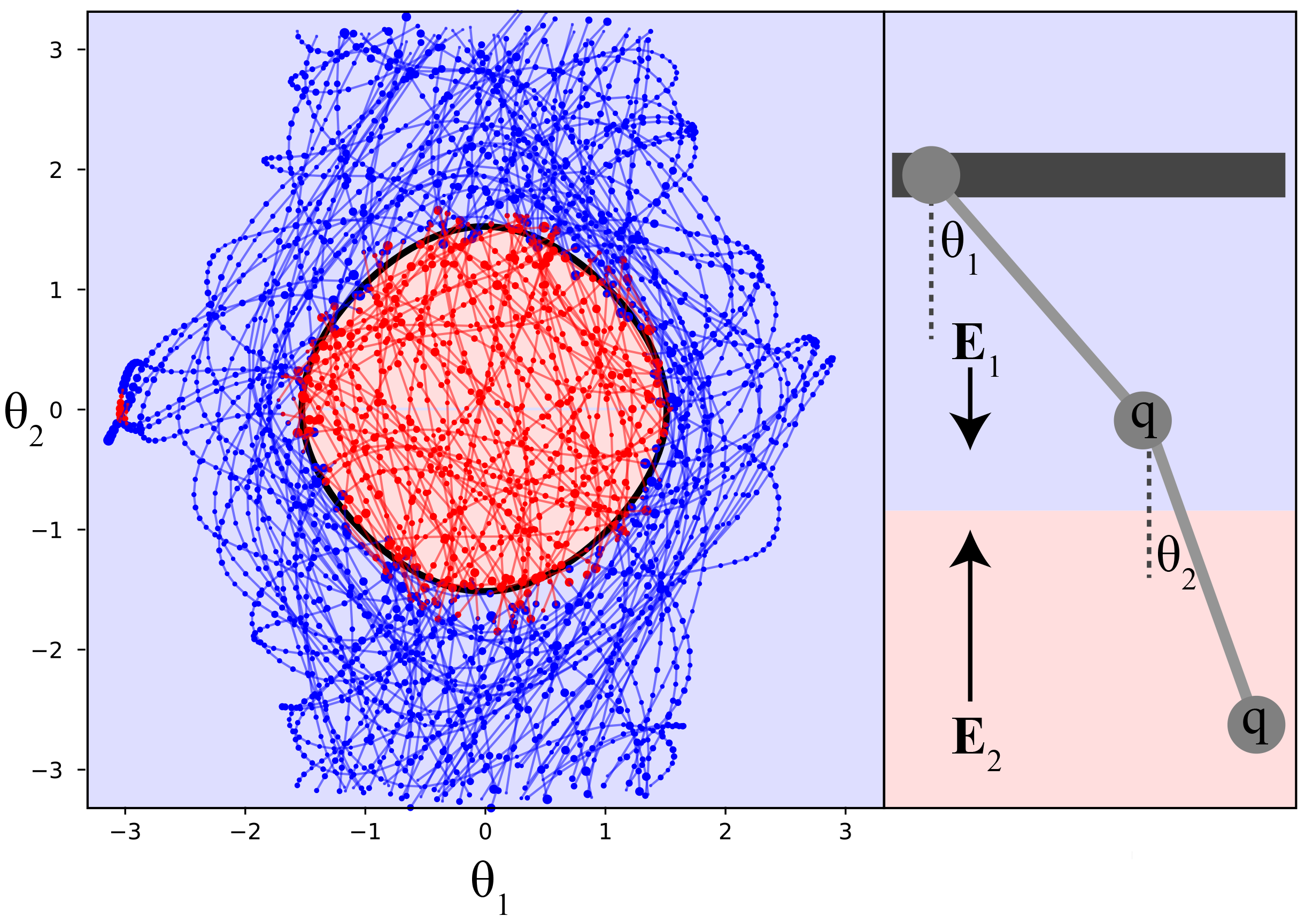}}
\vskip-5mm
\caption{In this mystery, a charged double pendulum moves through two different electric fields $\E_1$ and $\E_2$, with a domain boundary corresponding to $\cos\theta_1+\cos\theta_2=1.05$ (the black curve above left, where the lower charge crosses the $\E$-field boundary).
The color of each dot represents the domain into which it is classified by a Newborn agent, and its area represents the description length of the error with which its position is predicted, for a precision floor $\epsilon\approx 0.006$. In this world, the Newborn agent has a domain prediction accuracy of 96.5\%.
\label{PendulumFig}
}
\end{figure}

Our double-pendulum mysteries (Appendix~\ref{appendix:double_pendulum}) are more challenging for all the agents, because the motion is more nonlinear and indeed chaotic.
Although none of our double-pendulum mysteries get exactly solved according to our very stringent above-mentioned criterion, \fig{PendulumFig}
 illustrates that the Newborn agent does a good job: it discovers the two domains and classifies points into them with an accuracy of 96.5\%. Overall, the Newborn agent has a median best accuracy of 91.0\% compared with the baseline of  76.9\%.
 The MSE prediction error is comparable to the baseline performance ($\sim4\times 10^{-4})$ in the median, since both architectures have similar large capacity.
 We analyze this challenge and opportunities for improvement below.

\section{Conclusions}

We have presented a toy ``AI Physicist" unsupervised learning agent 
centered around the learning and manipulation of theories, which in polynomial time learns to parsimoniously predict both aspects of the future (from past observations) and the domain in which these predictions are accurate.

\subsection{Key findings}

Testing it on a suite of mystery worlds involving random combinations of gravity, electromagnetism, harmonic motion and elastic bounces, we found that its divide-and-conquer and Occam's razor strategies effectively identified domains with different laws of motion and reduced the mean-squared prediction error billionfold, typically recovering integer and rational theory parameters exactly. 
These two strategies both encouraged prediction functions to specialize: 
the former on the domains they handled best, and the latter on the data points within their domain that they handled best.
Adding the lifelong learning strategy 
greatly accelerated learning in novel environments.

\subsection{What has been learned?}

Returning to the broader context of unsupervised learning from \Sec{IntroSec} raises two important questions: what is the {\it difficulty} of the problems that our AI physicist solved, and what is the {\it generality} of our paradigm?

In terms of {\it difficulty}, our solved physics problems are clearly on the easier part of the spectrum, so if we were to have faced the {\it supervised} learning problem where the different domains were pre-labeled, the domain learning would have been a straightforward classification task and the forecasting task could have been easily solved by a standard feedforward neural network.
Because the real world is generally unlabeled, we instead tackled the more difficult problem where boundaries of multiple domains had to be learned concurrently with the dynamical evolution rules in a fully 
unsupervised fashion. 
The dramatic performance improvement over a traditional neural network seen in Table~\ref{DetailedResultsTable} reflects the power of the divide-and-conquer and Occam's razor strategies, and their robustness is indicated by the the fact that unsupervised domain discovery worked well even for the two-field non-linear double-pendulum system whose dynamic is notoriously chaotic and whose domain boundary is the curved rhomboid $\cos\theta_1 + \cos\theta_2 = 1.05$.

In terms of {\it generality}, our core contribution lies in the AI physicist paradigm we propose (combining divide-and-conquer, Occam’s razor, unification and lifelong learning), not in the specific implementation details.
Here we draw inspiration from the history of the Turing Machine: Turing's initial implementation of a universal computer was very inefficient for all but toy problems, but his framework laid out the essential architectural components that subsequent researchers developed into today's powerful computers.
What has been learned is that our AI physicist paradigm outperforms traditional deep learning
on a test suite of problems even though it is a fully general paradigm that is was not designed specifically for these problems. For example, it is defined to work for an arbitrary number of input spatial dimensions, spatial domains, past time steps used, boundaries of arbitrary shapes, and evolution laws of arbitrary complexity.

From the above-mentioned successes and failures of our paradigm, we have also learned about promising opportunities for improvement of the implementation which we will now discuss.
First of all, the more modest success in the double-pendulum experiments illustrated the value of  learned theories being {\it simple}: if they are highly complex, they are less likely to unify or generalize to future environments, and the correspondingly complex baseline model will have less incentive to specialize because it has enough expressive power to approximate the motion in all domains at once. 
It will therefore be valuable to improve techniques for simplifying complex learned neural nets.
The specific implementation details for the Occam's Razor toolkit would then change, but the principle and numerical objective would remain the same: reducing their total description length from \eq{description_length}.
There are many promising opportunities for this using techniques from the Monte-Carlo-Markov-Chain-based and genetic techniques \cite{real2017large}, reinforcement learning \cite{zoph2016neural,baker2016designing} and symbolic regression \cite{schmidt2009distilling,udrescu2019ai} 
literature to simplify and shrink the model architecture.
Also, it will be valuable and straightforward to generalize our implementation to simplify not only the prediction functions, but also the classifiers, for example to find sharp domain boundaries composed of hyperplanes or other simple surfaces.

Analogously, there are many ways in which the unification and life-long learning toolkits can be improved while staying within our AI physicist paradigm. For example, unification can undoubtedly be improved by using more sophisticated clustering techniques for grouping the learned theories with similar ones. Life-long learning can probably be made more efficient by using better methods for determining which previous theories to try when faced by new data, for example by training a separate neural network to perform this prediction task.

\subsection{Outlook}
In summary, these and other improvements to the algorithms that implement our AI Physicist paradigm could enable future unsupervised learning agents to learn simpler and more accurate models faster from fewer examples, and also to discover accurate symbolic expressions for more complicated physical systems. More broadly, AI has been used with great success to tackle problems in diverse areas of physics,
ranging from quantum state reconstruction \cite{carrasquilla2019reconstructing} to phase transitions \cite{carrasquilla2017machine,wang2016discovering,van2017learning}, planetary dynamics \cite{lam2018machine} and particle physics \cite{baldi2014searching}. 
We hope that building on the ideas of this paper may
one day enable AI to help us discover entirely novel physical theories from data.

 % - We've limited this to finding difference equations. Point out that trivially convertible to differential equations for linear domains. Searching for differential equations that predict our solutions is a harder problem warranting further work for nonlinear domains. 
  
 % - Mention the pixelized image challenge? Or better not to give away our idea cheaply and instead do separate paper on it?

  %* Time discretization. Way simpler with continuous time, since it avoids annoying edge cases where two successive steps are in different domains. Sorting this out during the learning stage, before the domains are known, requires additional domains and sending on THREE rather than TWO past data points, with associated degeneracies.

{\bf Acknowledgements:} 
This work was supported by the The Casey and Family Foundation, the Ethics and Governance of AI Fund, the Foundational Questions Institute and the Rothberg Family Fund for Cognitive Science. We thank Isaac Chuang, John Peurifoy and Marin Solja\v{c}i\'c for helpful discussions and suggestions,  and the Center for Brains, Minds, and Machines (CBMM) for hospitality.

\appendix

\section{AI Physicist Algorithm}
\label{AIphysicistAlgo}

The detailed AI Physicist algorithm is presented in Algorithm~\ref{alg:overall_algorithm}, with links to each of the individual sub-algorithms.
Like most numerical methods, the algorithm contains a number of hyperparameters that can be tuned to optimize performance; Table~\ref{hyperparameter_table} lists them and their settings for our numerical experiments.

\begin{algorithm}[h]
\caption{\textbf{AI Physicist: Overall algorithm}}
\label{alg:overall_algorithm}
\begin{algorithmic}
\STATE Given observations $D=\{(\x_t, \y_t)\}$ from new environment:
\STATE 1: $\mathbfcal{T}_{M_0}\gets \text{\textbf{Hub}.propose-theories}(D, M_0)$ (Alg.~\ref{alg:theory_propose})
\STATE 2: $\mathbfcal{T}\gets\text{differentiable-divide-and-conquer}(D, \mathbfcal{T}_{M_0})$(Alg.~\ref{alg:divide_and_conquer})
\STATE 3: \textbf{Hub}.add-theories($\mathbfcal{T},D$) (Alg.~\ref{alg:add_theory})
\STATE
\STATE Organizing theory hub:
\STATE $\mathbfcal{T}\gets$\textbf{Hub}.Occam's-Razor-with-MDL($\mathbfcal{T},D$) (Alg.~\ref{alg:MDL_simplification})
\STATE $\mathbfcal{T}\gets$\textbf{Hub}.unify($\mathbfcal{T}$) (Alg.~\ref{alg:unification})
\end{algorithmic}
\end{algorithm}

\section{The Differentiable Divide-and-Conquer (DDAC) Algorithm}
\label{DivideAndConquerAlgo}

\begin{algorithm}[t]
\caption{\textbf{AI Physicist: Differentiable Divide-and-Conquer with Harmonic Loss}}
\label{alg:divide_and_conquer}
\begin{algorithmic}
   \STATE {\bfseries Require} Dataset $\D=\{(\x_t, \y_t)\}$
   \STATE {\bfseries Require $M$}: number of initial total theories for training
   \STATE {\bfseries Require $\mathbfcal{T}_{M_0}=\{(\f_i,c_i)\}, i=1,...,M_0, 0\leq M_0\leq M$}: theories proposed from theory hub
   \STATE {\bfseries Require $K$}: number of gradient iterations
   \STATE {\bfseries Require $\beta_\f, \beta_\c$}: learning rates
   \STATE {\bfseries Require $\epsilon_0$}: initial precision floor
   \STATE
   \STATE 1: Randomly initialize $M-M_0$ theories $\T_i, i=M_0+1,$
   \STATE \ \ \ \ $...,M$. Denote $\mathbfcal{T}=(\T_1,,...,\T_M)$, $\mathbf{f_\vtheta}=(\f_1,...,\f_M)$,
   \STATE \ \ \ \ $\mathbf{c_\vphi}=(c_1,...c_M)$ with learnable parameters $\vtheta$ and $\vphi$.
  \STATE
  \STATE // \textit{Harmonic training with DL loss:}
   \STATE 2: $\epsilon\gets\epsilon_0$
   \STATE 3: \textbf{for} $\mathit{k}$ \textbf{in} $\{1,2,3,4,5\}$ \textbf{do}:
   \STATE 4: \ \ \ \ $\mathbfcal{T}\gets \textbf{IterativeTrain}(\mathbfcal{T}, \D, \ell_{\DL, \epsilon}, \Ell_{-1})$, where
   \STATE \ \ \ \ \ \ \ \  $\Ell_{-1}\equiv\sum_t\left(\frac{1}{M}\sum_{i=1}^M \ell[\f_i(\x_t), \y_t]^{-1}\right)^{-1}$ (Eq. \ref{generalized_mean_risk_empirical})
 \STATE 5: \ \ \ \  $\epsilon\gets\text{set\_epsilon}(\mathbfcal{T},\D)$ //
   \textit{median prediction error}
    \STATE 6: \textbf{end for}
  \STATE
   \STATE // \textit{Fine-tune each theory and its domain:}
   \STATE 7: \textbf{for} $\mathit{k}$ \textbf{in} $\{1,2\}$ \textbf{do}:
   \STATE 8: \ \ \ \ $\mathbfcal{T}\gets \textbf{IterativeTrain}(\mathbfcal{T}, \D, \ell_{\DL, \epsilon}, \Ell_\text{dom})$, where
   \STATE \ \ \ \ \ \ \ \ $\Ell_{\text{dom}}\equiv\sum_t \ell[\f_{i_t}(\x_t), \y_t]$ with $i_t=\text{arg max}_i\c_i(\x_t)$ 
 \STATE 9: \ \ \ \  $\epsilon\gets\text{set\_epsilon}(\mathbfcal{T},\D)$ //
   \textit{median prediction error}
    \STATE 10: \textbf{end for}
   \STATE 11: \textbf{return} $\mathbfcal{T}$
   \STATE
   \STATE \textbf{subroutine IterativeTrain}$(\mathbfcal{T}, \D, \ell, \Ell):$
   
   \STATE s1: \textbf{for} $k$ in $\{1,...,K\}$ \textbf{do}:
   \STATE \textit{\ \ \ \ \ \ \ //   Gradient descent on $\mathbf{f_\vtheta}$ with loss $\Ell$:}
   \STATE s2: \ \ \ $\g_\f\gets\nabla_\vtheta\Ell[\mathbfcal{T},D,\ell]$
   \STATE s3:\ \ \ \ \ Update $\vtheta$ using gradients $\mathbf{g}_\f$ (e.g. Adam \cite{kingma2014adam} or SGD)
   \STATE \textit{\ \ \ \ \ \ \ // Gradient descent on $\c_\vphi$ with the best performing}
   \STATE \textit{\ \ \ \ \ \ \ \ \ \ \ theory index as target:}
   \STATE s4: \ \ \ $b_t\gets \argmin_i\{\ell[\f_i(\x_t), \y_t]\}, \forall t$
   \STATE s5: \ \ \ \ $\g_\c\gets \nabla_\vphi\sum_{(\x_t, \cdot)\in \D}\text{CrossEntropy}[\text{softmax}(\c_\vphi(\x_t)), b_t]$
   \STATE s6:\ \ \ \ \ Update $\vphi$ using gradients $\g_\c$ (e.g. Adam \cite{kingma2014adam} or SGD)
   \STATE s7: \textbf{end for}
   \STATE s8: $\mathbfcal{T}\gets \text{AddTheories}(\mathbfcal{T},D,\ell,\Ell)$\ //Optional
   \STATE s9: $\mathbfcal{T}\gets \text{DeleteTheories}(\mathbfcal{T},D,\ell)$\ //Optional
   \STATE s10: \textbf{return} $\mathbfcal{T}$
\end{algorithmic}
\end{algorithm}

Here we elaborate on our differentiable divide-and-conquer (DDAC) algorithm with generalized-mean loss $\Ell_\gamma$ (Eq. (\ref{generalized_mean_risk_empirical})). This loss with $\gamma <0$ works with a broad range of distance functions $\l$ satisfying Theorem~\ref{thm:theorem_gradient}. Since the goal of our AI Physicist is to minimize the overall description length (DL) from \eq{description_length}, we choose $\l$ to be the DL loss function of \eq{RealDLeq} together with $\gamma=-1$ (harmonic loss), which works quite well in practice.

Algorithm~\ref{alg:divide_and_conquer} describes our differentiable divide-and-conquer implementation, which consists of two stages. In the first stage (steps 2-6), it applies the subroutine $\text{IterativeTrain}(\mathbfcal{T},D,\ell_{\DL,\epsilon},\Ell_{-1})$ with harmonic loss $\Ell_{-1}$ to train the theories $\mathbfcal{T}$ a few times with the precision floor $\epsilon$ gradually lowered according to the following annealing schedule. We set the initial precision floor $\epsilon$ to be quite large so that $\ell$ initially approximates an MSE loss function.
After each successive iteration, we reset $\epsilon$ to the median prediction error.

The DL loss function from \eq{RealDLeq} is theoretically desirable but tricky to train,  both because it is non-convex and because it is quite flat and uninformative far from its minimum. 
Our annealing schedule helps overcome both problems: initially when $\epsilon$ is large, it approximates MSE-loss which is convex and guides the training to a good approximate minimum, which further training accurately pinpoints as $\epsilon$ is reduced.

The subroutine $\text{IterativeTrain}$ forms the core of the algorithm. In the first stage (steps 2-6), it uses the harmonic mean of the DL-loss of multiple prediction functions $\f_\vtheta=(\f_1,...,\f_M)$
(\ie, \eq{generalized_mean_risk_empirical} with $\gamma=-1$ and $\ell=$DL)
to simultaneously train these functions, encouraging them to each specialize in the domains where they predict best (as proven by Theorem \ref{thm:theorem_gradient}), and simultaneously trains the domain classifier $\c_\vphi=(c_1,...,c_M)$ using each example's best-performing prediction function as target, with categorical cross-entropy loss. 
After several rounds of $\text{IterativeTrain}$ with successively lower precision floors, each prediction function typically becomes good at predicting part of the dataset, and the domain classifier becomes good at predicting for each example which prediction function will predict best. 

AddTheories($\mathbfcal{T}, D,\ell,\Ell$) inspects each theory $\mathcal{T}_i$ describing at least a large fraction $\eta_{\rm insp}$ (we use $30\%$) of the examples to see if a non-negligible proportion $\eta_{\rm split}$ of examples (we use a threshold of $5\%$) of the examples inside its domain have MSE larger than a certain limit $\epsilon_{\rm add}$ (we use $2\times10^{-6}$) and thus warrant splitting off into a separate domain.

If so, it uses those examples to initialize a new theory $\mathcal{T}_{M+1}$, and performs tentative training together with other theories using IterativeTrain without steps s8 and s9 (it is also possible to allow steps s8 and s9 in this recursive calling of IterativeTrain, which will enable a recursive adding of theories for not-well-explained data, and may enable a more powerful DDAC algorithm). If the resulting loss $\Ell$ is smaller than before adding the new theory, $\mathcal{T}_{M+1}$ is accepted and retained, otherwise it is rejected and training reverts to the checkpoint before adding the theory. 
DeleteTheories($\mathbfcal{T},D,\ell$) deletes theories whose domain or best-predicted examples cover a negligible fraction of the examples (we use a delete threshold $\eta_{\rm del}=0.5\%$).

In the second stage (steps 7-10), the IterativeTrain is applied again, but the loss for each example $(\x_t,\y_t)$ is using only the theory that the domain classifier $\c_\vphi=(c_1,c_2,...c_M)$ predicts (having the largest logit). In this way, we iteratively fine-tune the prediction functions $\{\f_i\}$ w.r.t. each of its domain, and fine-tune the domain to the best performing theory at each point.
The reason that we assign examples to domains using our 
domain classifier rather than prediction accuracy is that the trained domains are likely to be simpler and more contiguous, thus generalizing better to unseen examples than, \eg, the nearest neighbor algorithm.

We now specify the default hyperparameters used for Algorithm \ref{thm:theorem_gradient} in our experiments (unless otherwise specified). We set the initial total number of theories $M=4$, from which $M_0=2$ theories are proposed from the theory hub. The initial precision floor $\epsilon_0=10$ and the number of gradient iterations $K=10000$. We use the Adam \cite{kingma2014adam} optimizer with default parameters for the optimization of both the prediction function and the domain classifier. We randomly split each dataset $D$ into $D_{train}$ and $D_{test}$ with 4:1 ratio. The $D_{test}$ is used only for evaluation of performance. The batch size is set to min(2000, $|D_{train}|$). We set the initial learning rate $\beta_\f=5\times 10^{-3}$ for the prediction functions $\f_\vtheta$ and
 $\beta_\c=10^{-3}$ for the domain classifier $\c_\vphi$. We also use a learning rate scheduler that monitors the validation loss every 10 epochs, 
 and divides the learning rate by 10 if the validation loss has failed to decrease after 40 monitoring points and stops training early if there is no decrease after 200 epochs --- or if 
the entire MSE loss for all the theories in their respective domains drops below $10^{-12}$.

To the main harmonic loss $\Ell_\gamma$, we add two regularization terms. One is $L_1$ loss whose strength increases quadratically from 0 to $\epsilon_{L1}$ during the first 5000 epochs and remains constant thereafter.
The second regularization term is a very small MSE loss of strength $\epsilon_{MSE}$, to encourage the prediction functions to remain not too far away from the target outside their domain.

\section{Occam's Razor with MDL Algorithm}
\label{appendix:OccamsRazorAlgo}

Pushing on after the DDAC algorithm with harmonic loss that minimizes the $\sum_t \DL(\u_t)$ term in Eq. (\ref{description_length}), the AI Physicist then strives to minimize the $\DL(\mathbfcal{T})$ term, which can be decomposed as $\DL(\mathbfcal{T})=\DL(\mathbf{f}_\vtheta)+\DL(\c_\vphi)$, where $\mathbf{f}_\vtheta=(\f_1,...\f_M)$ and $\c_\vphi=(c_1,...c_M)$. We focus on minimizing $\DL(\mathbf{f}_\vtheta)$, since in different environments the prediction functions $\f_i$ can often be reused, while the domains may differ. 
As mentioned, we define $\DL(\f_\vtheta)$ simply as the sum of the description lengths of the numbers parameterizing $\f_\vtheta$:
\beq{DL_for_f}
\DL(\mathbf{f}_\vtheta)=\sum_j \DL(\theta_j).
\eeq
This means that 
$\DL(\mathbf{f}_\vtheta)$ can be significantly reduced if an irrational parameter is replaced by a simpler rational number.

\begin{algorithm}[t]
\caption{\textbf{AI Physicist: Occam's Razor with MDL}}
\label{alg:MDL_simplification}
\begin{algorithmic}
   \STATE {\bfseries Require} Dataset $\D=\{(\x_t, \y_t)\}$
   \STATE {\bfseries Require $\mathbfcal{T}=\{(\f_i,c_i)\}, i=1,...,M$}: theories trained after Alg.~\ref{alg:divide_and_conquer}
   \STATE {\bfseries Require $\epsilon$}: Precision floor for $\ell_{\DL,\epsilon}$
   \STATE 1: \textbf{for} $i$ in $\{1,...,M\}$ \textbf{do}:
  \STATE 2: \ \ \ \ \ $\D^{(i)}\gets\{(\x_t,\y_t)|\argmax_j\{c_j(\x_t)\}=i\}$
  \STATE 3:\ \ \ \ \ \  $\f_i\gets\textbf{MinimizeDL}(\text{collapseLayers},\f_i, D^{(i)},\epsilon)$
  \STATE 4:\ \ \ \ \ \   $\f_i\gets\textbf{MinimizeDL}(\text{localSnap},\f_i, D^{(i)},\epsilon)$
  \STATE 5:\ \ \ \ \ \  $\f_i\gets\textbf{MinimizeDL}(\text{integerSnap},\f_i, D^{(i)},\epsilon)$
  \STATE 6:\ \ \ \ \ \  $\f_i\gets\textbf{MinimizeDL}(\text{rationalSnap},\f_i, D^{(i)},\epsilon)$
  \STATE 7:\ \ \ \ \ \  $\f_i\gets\textbf{MinimizeDL}(\text{toSymbolic},\f_i, D^{(i)},\epsilon)$
  \STATE 8: \textbf{end for}
  \STATE 9: \textbf{return} $\mathbfcal{T}$
  \STATE
  \STATE \textbf{subroutine MinimizeDL}(transformation, $\f_i$, $D^{(i)}$,$\epsilon$):
  \STATE s1:\ \textbf{while} transformation.is\_applicable($\f_i$) \textbf{do}:
  \STATE s2:\ \ \ \ \ \  $\text{dl}_0\gets\DL(\f_i)+\sum_{(\x_t,\y_t)\in D^{(i)}}\ell_{\DL,\epsilon}[\f_i(\x_t), \y_t]$
  \STATE s3:\ \ \ \ \ \  $f_{\text{clone}}\gets \f_i$  // \textit{clone $\f_i$ 
  in case transformation fails}
  \STATE s4:\ \ \ \ \ \  $\f_i\gets \text{transformation}(\f_i)$
  \STATE s5:\ \ \ \ \ \  $\f_i\gets\text{Minimize}_{\f_i}\sum_{(\x_t, \y_t)\in \D^{(i)}}\ell_{\DL,\epsilon}[\f_i(\x_t),\y_t]$
  \STATE s6:\ \ \ \ \ \  $\text{dl}_1\gets\DL(\f_i)+\sum_{(\x_t,\y_t)\in D^{(i)}}\ell_{\DL,\epsilon}[\f_i(\x_t), \y_t]$ 
 % \STATE s8:\ \ \ \ \ \ \ \ \ \ \ $\f_i\gets f_{\text{clone}}$
\STATE s7:\ \ \ \ \ \  \textbf{if} $\text{dl}_1>\text{dl}_0$ $\textbf{return}$ $f_{\text{clone}}$
  \STATE s8:\ \textbf{end while}
  \STATE s9:\ \textbf{return} $\f_i$
\end{algorithmic}
\end{algorithm}

If a physics experiment or neural net training produces a parameter $p=1.999942$, it would be natural to interpret this as a hint, and to check if $p=2$ gives an equally acceptable fit to the data. We formalize this by replacing any real-valued parameter $p_i$ in our theory $\T$ by its nearest integer if this reduces the total description length in \eq{description_length}, as detailed below. We start this search for  integer candidates with the parameter that is closest to an integer, refitting for the other parameters after each successful ``integer snap".
 
 What if we instead observe a parameter $p=1.5000017$? Whereas generic real numbers have a closest integer, they lack a closest rational number. Moreover, as illustrated in \fig{RationalComplexityFig}, we care not only about closeness (to avoid increasing the second term in \eq{description_length}), but also about simplicity (to reduce the first term).  
  To rapidly find the best ``rational snap" candidates (dots in \fig{RationalComplexityFig} that lie both near $p$ and far down), we perform a continued fraction expansion of $p$ and use each series truncation as a rational candidate.    
We repeat this for all parameters in the theory $\T$, again accepting only those snaps that reduce the total description length. 
We again wish to try the most promising snap candidates first; to rapidly identify promising candidates without having to recompute the second term in \eq{description_length}, we evaluate all truncations of all parameters as in \fig{NumberMDLfig}, comparing the description length of the rational approximation $q=m/n$ with the description length of the approximation error $|p-q|$. The most promising candidate minimizes their sum, \ie, lies furthest down to the left of the diagonal in the figure. The figure illustrates how, given the parameter vector $\p=\{\pi,\sqrt{2},3.43180632382353\}$, the first snap to be attempted will replace the third parameter by $53/17$.

  % Our algorithm implements iterated DL-minimization is presented in Appendix \ref{appendix:OccamsRazorAlgo}.
We propose Algorithm \ref{alg:MDL_simplification} to implement the above minimization of $\DL(\f_\vtheta)$ without increasing $\DL(\mathbfcal{T},D)$ (Eq. \ref{description_length}). For each theory $\T_i=(\f_i, c_i)$, we first extract the examples $D^{(i)}$ inside its domain, then perform a series of tentative transformations (simplifications) of the prediction function $\f_i$ using the MinimizeDL subroutine. This subroutine takes $\f_i$, the transformation, and $D^{(i)}$ as inputs and repeatedly applies the transformation to $\f_i$. After each such transformation, it fine-tunes the fit of $\f_i$ to $D^{(i)}$ using gradient descent. For determining whether to accept the transformation, Algorithm \ref{alg:MDL_simplification} presents the simplest 0-step patience implementation: if the description length $\text{dl}=\DL(\f_i)+\sum_{(\x_t,\y_t)\in D^{(i)}}\ell_{\DL,\epsilon}(\f_i(\x_t), \y_t)$ for theory $i$ decreases, then apply the transformation again if possible, otherwise exit the loop. In general, to allow for temporary increase of DL during the transformations, a non-zero patience can be adopted: at each step, save the best performing model as the pivot model, and if DL does not decrease during $n$ consecutive transformations inside MinimizeDL, exit the loop. In our implementation, we use a 4-step patience.

We now detail the five transformations used in Algorithm \ref{alg:MDL_simplification}. The collapseLayer transformation finds all successive layers of a neural net where the lower layer has linear activation, and combines them into one. The toSymbolic transformation transforms $\f_i$ from the form of a neural net into a symbolic expression (in our implementation, from a PyTorch net to a SymPy symbolic lambda expression). These two transformations are one-time transformations (for example, once $\f_i$ has been transformed to a symbolic expression,  toSymbolic cannot be applied to it again.) The localSnap transformation successively sets the incoming weights in the first layer to 0, thus favoring inputs that are closer to the current time step. 
The integerSnap transformation finds the (non-snapped) parameters in $\f_i$ that is closest to an integer, and snaps it to that integer. The rationalSnap transformation finds the (non-snapped) parameter in $\f_i$ that has the lowest bit sum when replaced by a rational number, as described in section \ref{sec:Occams_Razor}, and snaps it to that rational number. The latter three transformations can be applied multiple times to $\f_i$, until there are no more parameters to snap in $\f_i$, or the transformation followed by fine-tuning fails to reduce the description length.

In the bigger picture, Algorithm~ \ref{alg:MDL_simplification} is an implementation of minimizing the $\text{DL}(\f_\vtheta)$ without increasing the total $\text{DL}(\mathbfcal{T},D)$, if the description length of $\f_\vtheta$ is given by Eq. (\ref{DL_for_f}). There can be other ways to encode $\mathbfcal{T}$ with a different formula for $\text{DL}(\mathbfcal{T})$, in which case the transformations for decreasing $\text{DL}(\mathbfcal{T})$ may be different. But the structure of the Algorithm \ref{alg:MDL_simplification} remains the same, with the goal of minimizing $\text{DL}(\f_\vtheta)$ without increasing $\text{DL}(\mathbfcal{T},D)$ w.r.t. whatever DL formula it is based on.

In the still bigger picture, Algorithm~ \ref{alg:MDL_simplification} is a computationally efficient approximate implementation of the MDL formalism, involving the following two approximations:
\begin{enumerate}
    \item The description lengths $\DL(x)$ for various types of numbers are approximate, for convenience. For example, the length of the shortest self-terminating bit-string encoding an arbitrary natural number $n$ grows slightly faster than our approximation $\log_2 n$, because self-termination requires storing not only the binary digits of the integer, but also the length of said bit string, recursively, requiring 
    $\log_2 n + \log_2\log_2 n + \log_2\log_2 n + ...$, where only the positive terms are included 
       \cite{rissanen1983universal}. Slight additional overhead is required to upgrade the encodings to actual programs in some suitable language, including encoding of whether bits encode integers, rational numbers, floating-point numbers, {\etc}. 
    \item If the above-mentioned $\DL(x)$-formulas were made exact, they would be mere {\it upper bounds} on the true minimum description length. For example, our algorithm gives a gigabyte description length for $\sqrt{2}$ with precision $\epsilon=256^{-10^9}$, even though it can be computed by a rather short program, and there is no simple algorithm for determining which numbers can be accurately approximated by algebraic numbers. Computing the true minimum description length is a famous numerically intractable problem.
\end{enumerate}

\section{Unification Algorithm}
\label{UnificationAlgo}

\begin{algorithm}[t]
\caption{\textbf{AI Physicist: Theory Unification}}
\label{alg:unification}
\begin{algorithmic}
\STATE {\bfseries Require \textbf{Hub}}: theory hub
\STATE {\bfseries Require $C$}: initial number of clusters
\STATE 1:\ \textbf{for} $(\f_i,c_i)$ \textbf{in} \textbf{Hub}.all-symbolic-theories \textbf{do}:
\STATE 2:\ \ \ \ \ \ $\text{dl}^{(i)}\gets \DL(\f_i)$
\STATE 3:\ \textbf{end for}
\STATE 4:\ $\{S_k\}\gets$Cluster $\{\f_i\}$ into $C$ clusters based on $\text{dl}^{(i)}$
\STATE 5:\ \textbf{for} $S_k$ \textbf{in} $\{S_k\}$ \textbf{do}:
\STATE 6:\ \ \ \ \ $(\g_{i_k},\h_{i_k})\gets\textbf{Canonicalize}(\f_{i_k})$, $\forall \f_{i_k}\in S_k$
\STATE 7:\ \ \ \ \ $\h_k^*\gets$ Mode of $\{\h_{i_k}|\f_{i_k}\in S_k\}$.
\STATE 8:\ \ \ \ \ $G_k\gets\{\g_{i_k}|\h_{i_k}=\h_k^*\}$
\STATE 9:\ \ \ \ \ \ \ $\g_{\p_k}\gets$Traverse all $\g_{i_k}\in G_k$ with synchronized steps, 
\STATE \ \ \ \ \ \ \ \ \ \ \ \ \ \ \ \ \ replacing the coefficient by a $\p_{j_k}$ when not all  
\STATE \ \ \ \ \ \ \ \ \ \ \ \ \ \ \ \ coefficients at the same position are identical.
\STATE 10:\ \ \ \ $\f_{\p_k}\gets \text{toPlainForm}(\g_{\p_k})$
\STATE 11:\ \textbf{end for}
\STATE 12:\ $\mathscr{T}\gets \{(\f_{\p_k},\cdot)\}$, $k=1,2,...C$
\STATE 13:\ $\mathscr{T}\gets\text{MergeSameForm}(\mathscr{T})$
\STATE 14:\ \textbf{return} $\mathscr{T}$

\STATE
\STATE \textbf{subroutine Canonicalize}($\f_i$):
\STATE s1:\ $\g_i\gets\text{ToTreeForm}(\f_i)$
\STATE s2:\ $\h_i\gets\text{Replace all non-input coefficient by a symbol $s$}$
\STATE \textbf{return} $(\g_i,\h_i)$

\end{algorithmic}
\end{algorithm}

\bigskip

The unification process takes as input the symbolic prediction functions $\{(\f_i,\cdot)\}$, and outputs master theories $\mathscr{T}=\{(\f_\p,\cdot)\}$ such that by varying each $\p$ in $\f_\p$, we can generate a continuum of prediction functions $\f_i$ within a certain class of prediction functions. The symbolic expression consists of 3 building blocks: operators (e.g. $+$,$-$,$\times$,$/$), input variables (e.g. $x_1, x_2$), and coefficients that can be either a rational number or irrational number. The unification algorithm first calculates the DL $\text{dl}^{(i)}$ of each prediction function, then clusters them into $K$ clusters using e.g. K-means clustering. Within each cluster $S_k$, it first canonicalizes each $\f_{i_k}\in S_k$ into a 2-tuple $(\g_{i_k},\h_{i_k})$, where $\g_{i_k}$ is a tree-form expression of $\f_{i_k}$ where each internal node is an operator, and each leaf is an input variable or a coefficient. When multiple orderings are equivalent (e.g. $x_1+x_2+x_3$ vs. $x_1+x_3+x_2$), it always uses a predefined partial ordering. $\h_{i_k}$ is the structure of $\g_{i_k}$ where all coefficients are replaced by an $s$ symbol. Then the algorithm obtains a set of $\g_{i_k}$ that has the same structure $\h_{i_k}$ with the largest cardinality (steps 7-8). This will eliminate some expressions within the cluster that might interfere with the following unification process. Step 9 is the core part, where it traverses each $\g_{i_k}\in G_k$ with synchronized steps using e.g. depth-first search or breath-first search. This is possible since each $\g_{i_k}\in G_k$ has the same tree structure $h_k^*$. During traversing, whenever encountering a coefficient and not all coefficients across $G_k$ at this position are the same, replace the coefficients by some symbol $\p_{j_k}$ that has not been used before. Essentially, we are turning all coefficients that varies across $G_k$ into a parameter, and the coefficients that do not vary stay as they are. In this way, we obtain a master prediction function $\f_{\p_k}$. Finally, at step 13, the algorithm merges the master prediction functions in $\mathscr{T}=\{(\f_{\p_k},\cdot)\}$ that have the exact same form, and return $\mathscr{T}$. The domain classifier is neglected during the unification process, since at different environments, each prediction function can have vastly different spacial domains. It is the prediction function (which characterizes the equation of motion) that is important for generalization.

\section{Adding and Proposing Theories}
\label{appendix:theory_proposal_adding}

Here we detail the algorithms adding theories to the hub and proposing them for use in new environments.
Alg.~\ref{alg:theory_propose} provides a simplest version of the theory proposing algorithm. Given a new dataset $D$, the theory hub inspects all theories $i$, and for each one, 
counts the number $n_i$ of data points
where it outperforms all other theories. The top $M_0$ theories with largest $n_i$ 
are then proposed. 

For theory adding after training with DDAC (Alg.~\ref{alg:divide_and_conquer}), each theory $i$ calculates its description length $\text{dl}^{(i)}$ inside its domain. If its $\text{dl}^{(i)}$ is smaller than a threshold $\eta$, then the theory $(\f_i,c_i)$ with its corresponding examples $D^{(i)}$ are added to the theory hub. The reason why the data $D^{(i)}$ are also added to the hub is that $D^{(i)}$ gives a reference for how the theory $(\f_i,c_i)$ was trained, and is also needed in the Occam's razor algorithm.

\begin{algorithm}[t]
\caption{\textbf{AI Physicist: Theory Proposing from Hub}}
\label{alg:theory_propose}
\begin{algorithmic}
\STATE {\bfseries Require \textbf{Hub}}: theory hub
\STATE {\bfseries Require} Dataset $\D=\{(\x_t, \y_t)\}$
\STATE {\bfseries Require $M_0$}: number of theories to propose from the hub

\STATE 1: $\{(\f_i,c_i)\}\gets \textbf{Hub}.\text{retrieve-all-theories()}$
\STATE 2: $D_{\text{best}}^{(i)}\gets\{(\x_t, \y_t)|\text{argmin}_j\ell_{\DL,\epsilon}[\f_j(\x_t), \y_t]=i\}$, $\forall i$ 
\STATE 3: $\mathbfcal{T}_{M_0}\gets\big\{(\f_i, c_i)\big| D_{\text{best}}^{(i)}\text{ ranks among } M_0\ \text{largest sets in}\ \{D_{\text{best}}^{(i)}\}\big\}$
\STATE 4:\ \textbf{return} $\mathbfcal{T}_{M_0}$
\end{algorithmic}
\end{algorithm}

\begin{algorithm}[t]
\caption{\textbf{AI Physicist: Adding Theories to Hub}}
\label{alg:add_theory}
\begin{algorithmic}
\STATE {\bfseries Require \textbf{Hub}}: theory hub
\STATE {\bfseries Require $\mathbfcal{T}=\{(\f_i,c_i\}$}: Trained theories from Alg.~\ref{alg:divide_and_conquer}
\STATE {\bfseries Require} Dataset $\D=\{(\x_t, \y_t)\}$
\STATE {\bfseries Require $\eta$}: DL threshold for adding theories to hub
\STATE 1: $D^{(i)}\gets\{(\x_t,\y_t)|\argmax_j\{c_j(\x_t)\}=i\}, \forall i$
\STATE 2: $\text{dl}^{(i)}\gets\frac{1}{|D^{(i)}|}\sum_{(\x_t,\y_t)\in D^{(i)}} \ell_{\DL,\epsilon}(\f_i(\x_t),\y_t), \forall i$
\STATE 3: \textbf{for} $i$ \textbf{in} $\{1,2,...|\mathbfcal{T}|\}$ \textbf{do}:
\STATE 4: \ \ \ \  \textbf{if} $\text{dl}^{(i)}<\eta$ \textbf{do}
\STATE 5: \ \ \ \ \ \ \ \ \  \textbf{Hub}.addIndividualTheory$((\f_i, c_i), D^{(i)})$ 
\STATE 6: \ \ \ \ \textbf{end if}
\STATE 7: \textbf{end for}
\end{algorithmic}
\end{algorithm}

\section{Time complexity}
\label{ComplexitySec}

\def\ndom{n_{\rm dom}}
\def\ndata{n_{\rm data}}
\def\ntheo{n_{\rm theo}}
\def\npar{n_{\rm par}}
\def\nmyst{n_{\rm myst}}

In this appendix, we list crude estimates of the time complexity of our AI physicist algorithm, \ie, of how its runtime scales with key parameters.

 \textbf{DDAC}, the differentiable divide-and-conquer algorithm, algorithm (Alg.~\ref{alg:divide_and_conquer}), is run once for each of the $\nmyst$ different mystery worlds,  with a total runtime scaling roughly as
$$\mathcal{O}(\nmyst\npar\ndata\ndom^2),$$
where  
$\npar$ is the average number of neural-network parameter in a theory, 
$\ndata$ is the average number of data points (time steps) per mystery
and
$\ndom$ is the number of discovered domains (in our case $\le 4$). The power of two for $\ndom$ appears because the time to evaluate the loss function 
scales as $\ndom$, and we need to perform of  order $\ndom$ training cycles to add the right number of theories. The $\npar$ scaling is due to that the forward and backward propagation of neural net involves successive matrix multiplied by a vector, which scales as $O(n^2)$ where $n\simeq \sqrt{\npar/N_\text{lay}^f}$ is the matrix dimension for each layer
and $N_\text{lay}^f$ is the number of layers.
Accumulating all layers we have $N_\text{lay}^f n^2=\npar$. We make no attempt to model how the number of training epochs needed to attain the desired accuracy depends on parameters.

Our \textbf{Lifelong learning} algorithm is also run once per mystery, with a time cost dominated by that for proposing new theories (Alg.~\ref{alg:theory_propose}), which scales as
$$\mathcal{O}(\nmyst\ndata\ntheo).$$
Here $\ntheo$ is the number of theories in theory hub.

In contrast, our {\bf Occam's Razor} algorithm (Alg. \ref{alg:MDL_simplification}) and {\bf unification} algorithm (Alg.~\ref{alg:unification}) are run once per learned theory, not once per mystery.
For Occam's Razor, the total runtime is dominated by that for snapping to rational numbers, 
which scales as
$$\mathcal{O}(\npar\ndata\ntheo).$$
For the unification, the total runtime scales as 
$\mathcal{O}(\npar\ntheo)$, which can be neglected relative to the cost of Occam's razor.

We note that all these algorithms have merely polynomial time complexity. 
The DDAC algorithm dominates the time cost; our mystery worlds were typically solved in about 1 hour each on a single CPU.
If vast amounts of data are available, it may suffice to analyze a random subset of much smaller size.

\section{Proof of Theorem 1 and Corollary}
\label{proof_theorem_gradient}
Here we give the proof for Theorem \ref{thm:theorem_gradient}, restated here for convenience.

\textbf{Theorem 1}\ 
\textit{Let $\hat{\y}^{(i)}_t\equiv \f_i(\x_t)$ denote the prediction of the target $\y_t$ by the function $\f_i$, $i=1,2,...M$.
Suppose that $\gamma<0$ and $\ell(\hat{\y}_t, \y_t) = \ell(|\hat{\y}_t - \y_t|)$ for a monotonically increasing function 
$\ell(u)$ that vanishes on $[0,u_0]$ for some $u_0\geq 0$, with $\ell(u)^\gamma$ differentiable and strictly convex for $u>u_0$. \\
Then if $0<\ell(\hat{\y}_t^{(i)}, \y_t) < \ell(\hat{\y}_t^{(j)}, \y_t)$, we have 
\beq{gradient_greater_appendix}
\left|\frac{\partial\Ell_\gamma}{\partial u^{(i)}_t}\right| > \left|\frac{\partial \Ell_\gamma}{\partial u^{(j)}_t}\right|,
\eeq
 where $u^{(i)}_t\equiv|\hat{\y}_t^{(i)} - \y_t|$}.

\textit{Proof}.
Since $u^{(i)}_t\equiv|\hat{\y}_t^{(i)} - \y_t|$ and $\ell(\hat{\y}_t, \y_t) = \ell(|\hat{\y}_t - \y_t|)$, 
the generalized mean loss $L_\gamma$ as defined in Eq. \ref{gradient_greater} can be rewritten as 
\beq{L_gamma_loss}
\Ell_\gamma=\sum_t\left(\frac{1}{M}\sum_{k=1}^M \ell(u^{(k)}_t)^{\gamma}\right)^{\frac{1}{\gamma}},
\eeq
which implies that
\begin{equation*}
\begin{aligned}
\left|\frac{\partial \Ell_\gamma}{\partial u^{(i)}_t}\right|&=\left|\frac{1}{\gamma M}\left(\frac{1}{M}\sum_{k=1}^M \ell(u^{(k)}_t)^{\gamma}\right)^{\frac{1}{\gamma}-1}\>\frac{d\ell(u_t^{(i)})^{\gamma}}{du_t^{(i)}}\right|\\
&=
\frac{1}{|\gamma| M}\left(\frac{1}{M}\sum_{k=1}^M \ell(u^{(k)}_t)^{\gamma}\right)^{\frac{1}{\gamma}-1}\>\left|\frac{d\ell(u_t^{(i)})^{\gamma}}{du_t^{(i)}}\right|.
\end{aligned}
\end{equation*}
Since only the last factor depends on $i$, 
proving \eq{gradient_greater_appendix} 
is equivalent to proving that
\beq{ProofEq2}
\left|\frac{d \ell(u_t^{(i)})^{\gamma}}{d u_t^{(i)}}\right|>\left|\frac{d \ell(u_t^{(j)})^{\gamma}}{d u_t^{(j)}}\right|.
\eeq

Let us henceforth consider only the case $u>u_0$, since the conditions $\ell(u_t^{(j)})>\ell(u_t^{(i)})>0$ imply
$u_t^{(j)}>u_t^{(i)}>u_0$.
Since $\gamma<0$, $\ell(u)>0$ and
$\ell'(u)\geq 0$, we have
$\frac{d \ell(u)^{\gamma}}{d u}=\gamma \ell(u)^{\gamma-1}\ell'(u)\leq0$,
so that 
$\left|\frac{d \ell(u)^{\gamma}}{d u}\right|=-\frac{d \ell(u)^{\gamma}}{d u}$.
Because $\ell(u)^\gamma$ is differentiable and strictly convex, its derivative $\frac{d \ell(u)^{\gamma}}{d u}$ is monotonically increasing,
implying that $\left|\frac{d \ell(u)^{\gamma}}{d u}\right|=-\frac{d \ell(u)^{\gamma}}{d u}$ is monotonically decreasing. 
Thus  $\left|\frac{d\ell(u_1)^{\gamma}}{du_1}\right|>\left|\frac{d\ell(u_2)^{\gamma}}{du_2}\right|$
whenever $u_1<u_2$.
Setting $u_1=|\hat{\y}_t^{(i)}-\y_t|$ and $u_2=|\hat{\y}_t^{(j)}-\y_t|$ therefore implies \eq{ProofEq2}, which completes the proof.

\bigskip

The following corollary \ref{corollary:loss_satisfy} demonstrates that the theorem applies to several popular loss functions as well as our two description-length loss functions.
\begin{corollary}
\label{corollary:loss_satisfy}
Defining $u\equiv |\hat{\y}-\y|$, the following loss functions which depend only on $u$ satisfy the conditions for Theorem~\ref{thm:theorem_gradient}:
\begin{enumerate}
    % \item MSE loss: $l(\hat{\y}_t,\y_t)=u_t^2$
    % \item Mean absolute-error loss: $l(\hat{\y}_t,\y_t)=u_t$
    \item
    $\ell(u)=u^r$ for any $r>0$, which includes MSE loss ($r=2$) and mean-absolute-error loss ($r=1$).
    \item Huber loss: $\ell_\delta(u)=\begin{cases}
    \frac{1}{2}u^2, &  u\in[0,\delta]\\
    \delta(u-\frac{\delta}{2}),  & \text{otherwise},
    \end{cases}$\\
    where $\delta>0$.
    \item Description length loss\\
    $\ell_{\DL,\epsilon}(u)=\frac{1}{2}\log_2\left(1+\left(\frac{u}{\epsilon}\right)^2\right)$.
    \item Hard description length loss\\
     $\ell_{\text{DLhard},\epsilon}(u)=\log_2\max\left(1,\frac{u}{\epsilon}\right)$.
\end{enumerate}
\end{corollary}

\textit{Proof}.
We have $u_0=0$ for (1), (2), (3), and $u_0=\epsilon$ for (4). All  four functions $\l$ are monotonically increasing, satisfy $\ell(0)=0$ and are differentiable for $u>u_0$, so all that remains to be shown is that 
$\ell(u)^\gamma$ is strictly convex for $u>u_0$, \ie, that $\frac{d^2 \ell(u)^\gamma}{d u^2}>0$ when $u>u_0$.

(1) For $\ell(u)=u^r$ and $u>0$,
we have $\frac{d^2 \ell(u)^\gamma}{d u^2}=\gamma r(\gamma r -1)u^{\gamma r - 2}>0$, since
$\gamma<0$ and $r>0$ implies that  $\gamma r<0$ and $\gamma r-1<0$, so 
$\ell(u)^\gamma$ is strictly convex for $u>0$.

(2) The Huber loss $\ell_\delta(u)$ is continuous with a continuous derivative.
It satisfies
$\frac{d^2 \ell(u)^\gamma}{d u^2}>0$ both for $0<u<\delta$ and for $\delta<u$ according to the above proof of (1), since 
$\ell_\delta(u)$ is proportional to $\ell^r$ in these two intervals with $r=2$ and $r=1$, respectively.
At the transition point $u=\delta$, this second derivative is discontinuous, but takes positive value approaching both from the left and from the right, so  $\ell(u)^\gamma$ is strictly convex.
More generally, any function $\ell(u)$ built by smoothly connecting functions $\ell_i(u)$ in different intervals will satisfy our theorem if the functions $\ell_i(u)$ do.

 (3) Proving strict convexity of $\ell(u)^\gamma$ when $\ell$ is the description length loss $\ell_{\DL,\epsilon}(u)=\frac{1}{2}\text{log}_2\left[1+\left(\frac{u}{\epsilon}\right)^2\right]$
 is equivalent to proving it when $\ell(u)=\ro(u)\equiv \ln(1+u^2)$, since convexity is invariant under horizontal and vertical scaling.
We thus need to prove that  
$$\frac{d^2 \ro(u)^\gamma}{d u^2}
=-\frac{2\gamma[\ln(1+u^2)]^{\gamma-2}}{(1+u^2)^2 [2u^2(1-\gamma)+(u^2-1)\ln(1+u^2)]}$$ 
is positive when $u>0$.
The factor 
$\frac{-2\gamma[\ln(1+u^2)]^{\gamma-2}}{(1+u^2)^2}$ is always positive. The other factor 
$$2u^2(1-\gamma)+(u^2-1)\text{log}(1+u^2)>2u^2+(u^2-1)\text{log}(1+u^2),$$ 
since $\gamma<0$. Now we only have to prove that the function
$$\chi(u)\equiv2u^2+(u^2-1)\text{log}(1+u^2)>0$$ 
when $u>0$. We have $\chi(0)=0$ and 
$$\chi'(u)=2u\left[\frac{1+3u^2}{1+u^2}+\text{log}(1+u^2)\right]>0$$ 
when $u>0$. Therefore $\chi(u)=\chi(0)+\int_0^u\chi'(u')du'>0$ when $u>0$, which completes the proof that $\ell_{\DL,\epsilon}(u)^\gamma$ is strictly convex for $u>0$.

(4) For the hard description length loss $\ell_{\text{DLhard},\epsilon}(u)=\log_2\max\left(1,\frac{u_t}{\epsilon}\right)$, 
we have $u_0=\epsilon$. 
When $u>\epsilon$, we have 
$\ell'_{\text{DLhard},\epsilon}(u)=\frac{1}{u\text{ln}2}>0$ and 
$$\frac{d^2 \ell^\gamma_{\text{DLhard},\epsilon}(u)}{d u^2}=
{\gamma\over\ln 2}\left(-1+\gamma-\text{ln}\frac{u}{\epsilon}\right)\frac{\left(\ln{u\over\epsilon}\right)^{\gamma-2}}{u^2} .$$
For $u>\epsilon$, the factor $\frac{\left(\ln{u\over\epsilon}\right)^{\gamma-2}}{u^2}$ is always positive, as is the factor  $\gamma(-1+\gamma-\text{ln}\frac{u}{\epsilon})$, since $\gamma<0$.
$\ell^\gamma_{\text{DLhard},\epsilon}(u)$ is therefore strictly convex for $u>\epsilon$.

\section{Eliminating Transition Domains}
\label{DomainElimination}

In this appendix, we show how the only hard problem our AI Physicist need solve is to determine the laws of motion far from domain boundaries, because once this is done, the exact boundaries and transition regions can be determined automatically.

\begin{figure}[pbt]
\centerline{\includegraphics[width=88mm]{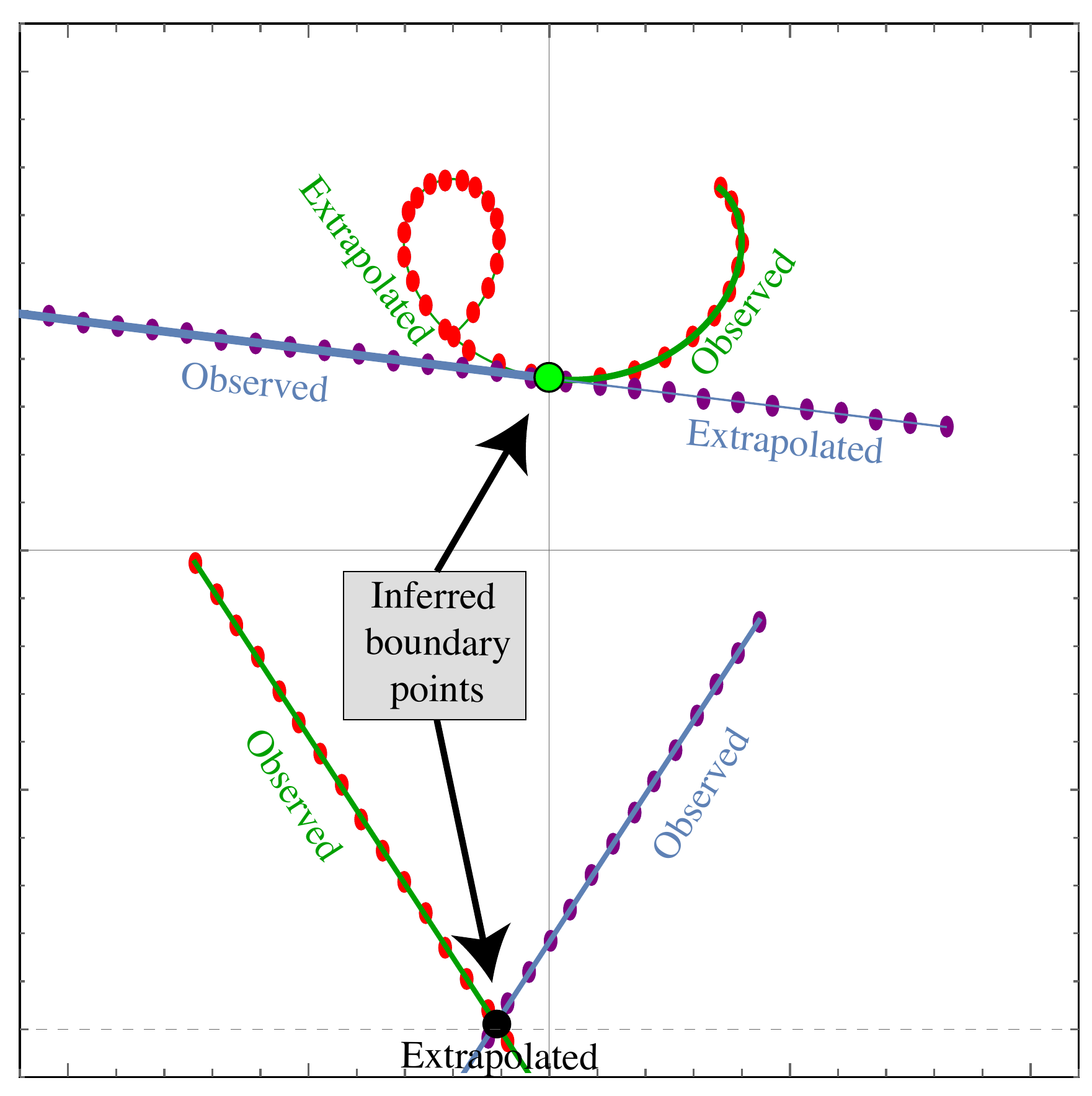}}
\caption{Points where forward and backward extrapolations agree (large black dots) are boundary points. The tangent vectors agree for region boundaries (upper example), but not for bounce boundaries (lower example). 
\label{BoundaryPointFig}
}
\end{figure}

Our AI Physicist tries to predict the next position vector $\y_t\in R^d$ from the
concatenation $\x_t =(\y_{t-T},...,\y_{t-1})$  of the last $T$ positions vectors. 
Consider the example shown in \fig{BoundaryPointFig}, where motion is predicted from the last $T=3$ positions in a space with $d=2$ dimensions containing $n=2$ domains with different physics (an electromagnetic field in the upper left quadrant and free motion elsewhere), as well as perfectly reflective boundaries. Although there are only two physics domains in the 2-dimensional space, there are many more types of domains in the $Td=6$-dimensional space of $\x_t$ from which the AI Physicist makes its predictions of $\y_t$. When a trajectory crosses the boundary between the two spatial regions, there can be instances where $\x_t$ contains 3, 2, 1 or 0 points in the first domain and correspondingly 0, 1, 2 or 3 points in the second domain. Similarly, when the ball bounces, there can be instances where $\x_t$ contains 3, 2, 1 or 0 points before the bounce and correspondingly 0, 1, 2 or 3 points after. Each of these situations involves a different function $\x_t\mapsto\y_t$ and a corresponding 6-dimensional domain of validity for the AI Physicist to learn.

Our numerical experiments showed that the AI Physicist typically solves the big domains (where all vectors in $\x_t$ lie in the same spatial region), but occasionally fails to find an accurate solution in some of the many small transition domains involving boundary crossings or bounces, where data is insufficient. Fortunately, simple post-processing can automatically eliminate these annoying transition domains with an algorithm that we will now describe.

The first step of the algorithm is illustrated in \fig{BoundaryPointFig}.
For each big domain where our AI Physicist has discovered the future-predicting function 
$\x_t\mapsto\y_t$, we determine the corresponding function that predicts the {\it past}
($\x_t\mapsto\y_{t-T-1}$) by fitting to forward trajectories generated with random initial conditions.
Now whenever a trajectory passes from a big domain through a transition region into another big domain, two different extrapolations can be performed: forward in time from the first big domain or backward in time from the second big domain. Using cubic spline interpolation, we fit continuous functions $\y_f(t)$ and $\y_b(t)$ (smooth curves in \fig{BoundaryPointFig}) to these forward-extrapolated and backward-extrapolated trajectories, and numerically find the time
\beq{BoundarpointEq}
t_*\equiv\argmin |\y_f(t)-\y_b(t)|
\eeq
when the distance between the two predicted ball positions is minimized.
If this minimum is numerically consistent with zero, so that $\y_f(t_*)\approx \y_b(t_*)$,
then we record this as being a boundary point.
If both extrapolations have the same derivative there, \ie, 
if $\y'_f(t_*)\approx \y'_b(t_*)$, then it is an interior boundary point between two different regions (\fig{BoundaryPointFig}, top), otherwise it is an external boundary point where the ball bounces (\fig{BoundaryPointFig}, bottom). 

\begin{figure}[phbt]
\centerline{\includegraphics[width=88mm]{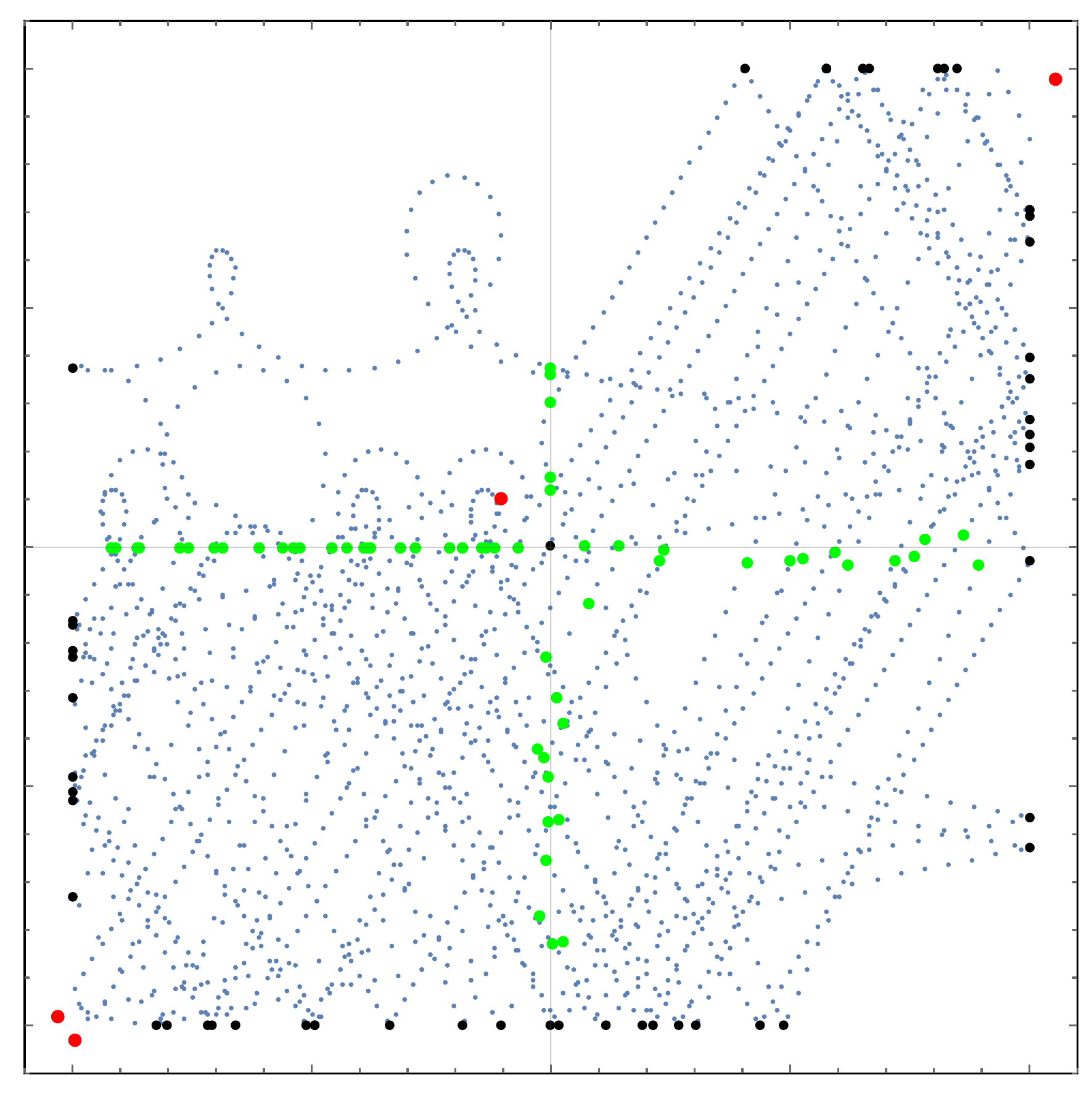}}
\caption{Example of automatically determined boundary points, for region boundary points (green), bounce boundary points (black) and failed cases (red).
\label{BoundaryPointsFig}
}
\end{figure}

\fig{BoundaryPointsFig} show these two types of automatically computed boundary points in green and black, respectively. These can now be used to retrain the domain classifiers to extend the big domains to their full extent, eliminating the transition regions.

Occasionally the boundary point determinations fill fail because of multiple transitions within $T$ time steps, \Fig{BoundaryPointsFig} illustrates that these failures (red dots) forces us to discard merely a tiny fraction of all cases, thus having a negligible affect on the ability to fit for the domain boundaries.

\section{Numerical Experiment Details}
\label{DetailedResultsSec}

In this appendix, we provide supplementary details on our benchmark problems.

\subsection{Mystery Worlds}
\label{appendix:mystery_worlds}

\textbf{World generation} Our mystery worlds consist of a ball elastically bouncing against the square boundary of the
two-dimensional spatial region where  $|x|\le 2$ and $|y|\le 2$ (see \fig{WorldExampleFig}).
In each of the four quadrants, one of the following laws of physics are selected, together with their parameters sampled from distributions as follows:
\begin{enumerate}
\item Free motion.
\item A uniform gravitational field ${\bf g}=(g_x,g_y,0)$ with $g_x, g_y$ drawn from a uniform distribution: $g_x, g_y\sim U[-5,5]$.
\item Harmonic motion with frequency $\omega$ around a line a distance $a$ from the origin, making an angle $\phi$ with the $x$-axis; 
$\omega\sim U[1,4]$, 
$a\sim U[0.2, 0.5]$,
$\phi\sim U[0, 2\pi]$.
\item A uniform electric field $\E=(E_x,E_y,0)$ and magnetic field $\B=(0,0,B_z)$; $E_x$, $E_y\sim U[-5,5]$, $B_z\sim U[0,10]$.
\end{enumerate}
To control the difficulty of the tasks and avoid near-degenerate scenarios, we keep only mystery worlds satisfying the following two criteria: (1) At least 0.01 separation between all equations of motion (EOM) in the same world, defined as the Euclidean distance between the vectors of coefficients specifying the EOM difference equations, and
(2)  at least 0.0015 of any non-integer parameter from its nearest integer.

\textbf{Trajectory simulation}
Within each world, we initialize the ball with a random position $(x, y)\sim U[-1,1]^2$ and velocity $(v_0\cos\theta_0, v_0 \sin\theta_0, 0)$; $v_0\sim U[0.1,0.5]$, $\theta_0\sim U[0,2\pi]$. We then compute its position for $N=4,000$ times steps $t=1,2,...,N$ with time interval 0.05.

Although the above-mentioned laws of physics are linear, the mapping
from past points $(\y_{t-T},...,\y_{t-1})$ to the next points $\y_t$ is generally non-linear because of region boundaries where the ball either bounces or transitions to a different physics region. An exception is when three successive points lie within the same region (with the same physics), which happens far from boundaries: in this case, 
the mapping from  $(\y_{t-2},\y_{t-1})\mapsto\y_t$ is deterministic and linear thanks to the differential equations of motion being second-order and linear.

\textbf{Architecture} For the Newborn and AI Physicist agents, each prediction function $\f_i$ is implemented as a $N_{lay}^f$-layer neural network with linear activation, with $N_{neur}^f$-neuron hidden layers (we use $N_{lay}^f=3$, $N_{neur}^f=8$ for our main experiments; see Table~\ref{hyperparameter_table}).
Each domain sub-classifier $c_i$ is implemented as a $N_{lay}^c$-layer neural net, with two hidden $N_{neur}^c$-neuron layers with leakyReLU activation
$\sigma(x)=\max\{0.3x,x\}$,
and the output layer having linear activation (we use $N_{lay}^c=3$, $N_{neur}^c=8$ for our main experiments). The baseline model is implemented as a single $N_{lay}^f$-layer neural net with two hidden 16-neuron layers with leakyReLU activation followed by a linear output layer.
Note that for a fair comparison, the baseline model has more hidden neurons, to roughly compensate for the Newborn and AI Physicist agents typically having multiple theories. The baseline network is nonlinear to boost its expressive power for  modeling the nonlinear prediction function of each world as a whole. For the domain classifier $\c=(c_1,c_2,...c_M)$, it is a $N_{lay}^c$-layer neural net where each hidden layer has $N_{neur}^c=8$ neurons and leakyReLU activation. The last layer has linear activation. See Table \ref{hyperparameter_table} for a list of hyperparameters.

\textbf{Evaluation} The unsupervised classification accuracy is defined as the fraction of correctly classified points, using the permutation of the learned domain labels that best matches the hidden ground truth domain labels. It is ``unsupervised" in the sense that there is no external supervision signal as to which domain label each point should be assigned to: the AI Physicist has to figure out the number of domains and their boundaries and assign each point to a domain, which is a difficult task.

We define a domain as {\it solved} if the agent discovers the its law of motion as difference equation 
 (prediction function) within the following stringent tolerance: all rational coefficients in the difference equation
 are exactly matched, and all irrational coeffients agree to an accuracy better than $10^{-4}$. 
Because of the nature of the physics problems, some of these difference equation coefficients take on the values $0$, $-1$, or $2$, so solving a region requires successful integer snapping as described in Section~\ref{sec:Occams_Razor}.
To make the problem even harder, we also fine-tune the magnetic field in five of the electromagnetic regions to make some of the coefficients simple fractions such as $1/3$ and $1/4$, thus making solving those regions contingent on successful rational snapping as described in Section~\ref{sec:Occams_Razor}. Domain solving can fail either by 
``undersnapping" (failing to approximate a floating-point number by a rational number) or
 `oversnapping" (mistakenly rounding to a rational number).
All our mystery worlds can be downloaded at \url{http://space.mit.edu/home/tegmark/aiphysicist.html}.

As shown in Appendix \ref{DomainElimination}, the only hard problem our AI Physicist or other algorithms need to solve is to determine the laws of motion away from domain boundaries. Therefore, we evaluate, tabulate and compare the performance of the algorithms only on interior points, \ie, excluding data points $(\x_t,\y_t)$ straddling a boundary encounter. %This applies to the MSE, unsupervised classification accuracy, and epochs to certain MSE level.

\begin{table*}
\begin{center}
{\footnotesize
\begin{tabular}{|l|rrr|rrr|rrr|rrr|}
\hline
&\multicolumn{3}{c|}{$\log_{10}$ MSE}
&\multicolumn{3}{c|}{Classification accuracy}
&\multicolumn{3}{c|}{Unsolved domains}
&\multicolumn{3}{c|}{Description length}\\
Regions&Base-&New-&AI&Base-&New-&AI&Base-&New-&AI&Base-&New-&AI\\
&line&born&phys&line&born&phys&line&born&phys&line&born&phys\\
\hline
Free + gravity		       &	      -4.12 &  -14.07 &      -14.08 &	    72.88\%&  100.00\%&      100.00\%&       2 &     0 &	 0 &		11310.4 &    59.4 &	   73.5 \\
Free + gravity		       &	      -4.21 &  -14.02 &      -14.04 &	    88.59\%&  100.00\%&      100.00\%&       2 &     0 &	 0 &		11271.5 &    60.3 &	   60.3 \\
Free + gravity		       &	      -3.69 &  -14.03 &      -14.03 &	    67.65\%&  100.00\%&      100.00\%&       2 &     0 &	 0 &		11364.2 &    60.2 &	   41.9 \\
Free + gravity		       &	      -4.18 &  -13.98 &      -13.98 &	    80.98\%&  100.00\%&      100.00\%&       2 &     0 &	 0 &		11341.7 &    60.6 &	   57.6 \\
Free + gravity		       &	      -4.51 &  -14.06 &      -14.07 &	    87.66\%&  100.00\%&      100.00\%&       2 &     0 &	 0 &		11289.3 &     5.2 &	   59.8 \\
Free + harmonic 	       &	      -3.77 &  -13.99 &      -13.94 &	    73.54\%&  100.00\%&      100.00\%&       2 &     0 &	 0 &		11333.8 &    94.4 &	  139.9 \\
Free + harmonic 	       &	      -3.60 &  -14.05 &      -13.89 &	    66.92\%&  100.00\%&      100.00\%&       2 &     0 &	 0 &		11337.4 &   173.0 &	  172.8 \\
Free + harmonic 	       &	      -3.77 &  -14.04 &      -13.95 &	    59.46\%&  100.00\%&      100.00\%&       2 &     0 &	 0 &		11317.5 &   156.0 &	  173.8 \\
Free + harmonic 	       &	      -5.32 &  -10.48 &      -13.14 &	    80.29\%&  100.00\%&      100.00\%&       2 &     1 &	 0 &		11219.5 &    91.6 &	   90.5 \\
Free + harmonic 	       &	      -3.64 &  -14.00 &      -13.89 &	    71.70\%&  100.00\%&      100.00\%&       2 &     0 &	 0 &		11369.6 &   143.7 &	  136.6 \\
Free + EM		       &	      -3.62 &  -13.95 &      -13.96 &	    82.77\%&  100.00\%&      100.00\%&       2 &     0 &	 0 &		11397.5 &   142.8 &	  284.9 \\
Free + EM		       &	      -4.13 &  -13.82 &      -13.67 &	    76.55\%&  100.00\%&      100.00\%&       2 &     0 &	 0 &		11283.0 &   306.2 &	  306.2 \\
Free + EM		       &	      -4.03 &  -13.45 &      -13.47 &	    74.56\%&   99.97\%&       99.97\%&       2 &     0 &	 0 &		11388.1 &   305.9 &	  307.9 \\
Free + EM		       &	      -4.31 &  -13.77 &      -13.62 &	    86.68\%&   99.91\%&       99.91\%&       2 &     0 &	 0 &		11257.7 &   152.0 &	  133.5 \\
Free + EM		       &	      -4.32 &  -14.00 &      -14.05 &	    84.55\%&  100.00\%&      100.00\%&       2 &     0 &	 0 &		11258.9 &   303.7 &	  303.8 \\
Free + EM rational	       &	      -3.45 &  -13.96 &      -13.95 &	    77.88\%&   99.96\%&       99.93\%&       2 &     0 &	 0 &		11414.9 &   194.2 &	  195.8 \\
Free + EM rational	       &	      -3.90 &  -13.96 &      -13.91 &	    71.13\%&  100.00\%&      100.00\%&       2 &     0 &	 0 &		11340.0 &   199.0 &	  199.0 \\
Free + EM rational	       &	      -4.12 &  -13.97 &      -13.90 &	    72.78\%&  100.00\%&      100.00\%&       2 &     0 &	 0 &		11330.7 &   198.8 &	  198.8 \\
Free + EM rational	       &	      -4.02 &  -14.07 &      -14.00 &	    77.92\%&  100.00\%&      100.00\%&       2 &     0 &	 0 &		11323.5 &   197.8 &	  197.8 \\
Free + EM rational	       &	      -4.83 &  -13.87 &      -13.86 &	    91.14\%&  100.00\%&      100.00\%&       2 &     0 &	 0 &		11247.1 &    10.3 &	   13.9 \\
Free + gravity + harmonic      &	      -4.08 &  -14.03 &      -13.95 &	    60.08\%&  100.00\%&      100.00\%&       3 &     0 &	 0 &		11269.0 &   191.8 &	  191.9 \\
Free + gravity + harmonic      &	      -4.31 &  -14.02 &      -13.66 &	    63.01\%&  100.00\%&      100.00\%&       3 &     0 &	 0 &		11334.2 &   170.4 &	   83.1 \\
Free + gravity + harmonic      &	      -4.01 &  -14.01 &      -13.99 &	    67.48\%&  100.00\%&      100.00\%&       3 &     0 &	 0 &		11351.0 &   168.7 &	  198.9 \\
Free + gravity + harmonic      &	      -3.64 &  -13.97 &      -13.88 &	    60.02\%&   99.97\%&       99.93\%&       3 &     0 &	 0 &		11374.6 &   225.7 &	  225.7 \\
Free + gravity + harmonic      &	      -4.11 &	-7.42 &       -7.43 &	    51.63\%&  100.00\%&       99.97\%&       3 &     1 &	 1 &		11313.7 &   193.5 &	  179.2 \\
Free + gravity + EM	       &	      -3.79 &  -13.93 &      -13.47 &	    57.89\%&  100.00\%&      100.00\%&       3 &     0 &	 0 &		11334.0 &   323.9 &	  346.8 \\
Free + gravity + EM	       &	      -4.18 &  -14.00 &      -14.00 &	    77.16\%&  100.00\%&      100.00\%&       3 &     1 &	 1 &		11301.0 &   277.9 &	   96.2 \\
Free + gravity + EM	       &	      -3.38 &  -13.58 &      -13.87 &	    53.33\%&  100.00\%&       99.96\%&       3 &     0 &	 0 &		11381.4 &   360.4 &	  364.0 \\
Free + gravity + EM	       &	      -3.46 &  -13.87 &      -13.89 &	    49.08\%&  100.00\%&      100.00\%&       3 &     0 &	 0 &		11370.1 &   354.0 &	  350.4 \\
Free + gravity + EM	       &	      -3.54 &  -13.69 &      -13.83 &	    51.28\%&  100.00\%&      100.00\%&       3 &     0 &	 0 &		11370.3 &   331.1 &	  320.7 \\
Free + harmonic + EM	       &	      -3.87 &  -13.82 &      -13.55 &	    67.27\%&  100.00\%&      100.00\%&       3 &     0 &	 0 &		11404.0 &   267.1 &	  275.4 \\
Free + harmonic + EM	       &	      -3.69 &  -13.87 &      -10.93 &	    56.02\%&   99.97\%&       99.94\%&       3 &     0 &	 0 &		11413.4 &   468.5 &	  464.9 \\
Free + harmonic + EM	       &	      -4.06 &  -13.39 &      -13.56 &	    70.87\%&  100.00\%&      100.00\%&       3 &     0 &	 0 &		11340.0 &   452.3 &	  452.3 \\
Free + harmonic + EM	       &	      -3.46 &  -13.94 &      -10.51 &	    59.02\%&   99.97\%&       99.93\%&       3 &     0 &	 0 &		11416.0 &   475.5 &	  471.9 \\
Free + harmonic + EM	       &	      -3.70 &  -13.75 &      -13.82 &	    61.67\%&  100.00\%&      100.00\%&       3 &     0 &	 0 &		11354.9 &   466.8 &	  466.8 \\
Free + gravity + harmonic + EM &	      -3.76 &  -13.82 &       -9.48 &	    27.93\%&  100.00\%&       99.94\%&       4 &     0 &	 0 &		11358.8 &   526.9 &	  530.4 \\
Free + gravity + harmonic + EM &	      -3.74 &  -13.00 &      -13.18 &	    40.80\%&  100.00\%&       99.97\%&       4 &     1 &	 1 &		11284.8 &   418.5 &	  389.1 \\
Free + gravity + harmonic + EM &	      -4.09 &  -13.97 &      -13.75 &	    35.69\%&  100.00\%&      100.00\%&       4 &     0 &	 0 &		11297.4 &   504.6 &	  504.6 \\
Free + gravity + harmonic + EM &	      -3.63 &  -13.80 &       -9.99 &	    31.61\%&  100.00\%&       99.97\%&       4 &     0 &	 0 &		11407.4 &   526.3 &	  526.2 \\
Free + gravity + harmonic + EM &	      -3.51 &	-6.37 &      -13.52 &	    32.97\%&  100.00\%&      100.00\%&       4 &     0 &	 0 &		11445.8 &   527.4 &	  527.5 \\
\hline
Median                         &	      -3.89 &  -13.95 &      -13.88 &	    67.56\%&  100.00\%&	     100.00\%&	    2.5&     0.00 &	  0.00 &	     11338.7 &   198.9 &       198.9 \\
Mean                           &	      -3.94 &  -13.44 &      -13.29 &	    65.51\%&   99.99\%&	      99.99\%&	    2.6&     0.10 &	  0.07 &	     11337.9 &   253.7 &       252.9 \\
\hline
\end{tabular}
}
\end{center}
\caption{Results for each of our first 40 mystery world benchmarks, as described in the section \ref{appendix:mystery_worlds}. Each number is the best out of ten trials with random initializations (using seeds 0, 30, 60, 90, 120, 150, 180, 210, 240, 270), and refers to big domains only. Based on the ``Unsolved domain" column, we count out of 40 worlds what's the percentage Baseline, Newborn and AI Physicist completely solve (has unsolved domain of 0), which goes to the ``Fraction of worlds solved" row in Table \ref{ResultsSummaryTable}.
\label{DetailedResultsTable}
}
\end{table*}

\begin{table*}
\begin{center}
{\footnotesize

\begin{tabular}{|l|rrr|rrr|rrr|rrr|}
\hline
&\multicolumn{3}{c|}{Epochs to $10^{-2}$}
&\multicolumn{3}{c|}{Epochs to $10^{-4}$}
&\multicolumn{3}{c|}{Epochs to $10^{-6}$}
&\multicolumn{3}{c|}{Epochs to $10^{-8}$}\\
Regions&Base-&New-&AI-&Base-&New-&AI&Base-&New-&AI&Base-&New-&AI\\
&line&born&phys&line&born&physi&line&born&phys&line&born&phys\\
\hline
Free+gravity                   &                100 &      85 &          85 &                 8440 &     120 &         120 &                $\infty$ &    4175 &        3625 &             $\infty$ &    6315 &        4890 \\
Free+gravity                   &                100 &      70 &          10 &                 4680 &     190 &          35 &                $\infty$ &    2900 &        4650 &             $\infty$ &    2995 &        6500 \\
Free+gravity                   &                 85 &     100 &          15 &                  $\infty$ &     135 &          30 &                $\infty$ &    8205 &        3815 &             $\infty$ &    9620 &        6455 \\
Free+gravity                   &                 95 &      75 &          20 &                 7495 &     140 &          25 &                $\infty$ &    6735 &        1785 &             $\infty$ &    8040 &        2860 \\
Free+gravity                   &                110 &      75 &           0 &                 1770 &     295 &          35 &                $\infty$ &    3740 &        3240 &             $\infty$ &    7030 &        3460 \\
Free + harmonic                &                 80 &      75 &          20 &                  $\infty$ &     145 &          25 &                $\infty$ &    2725 &        4050 &             $\infty$ &    2830 &        6145 \\
Free + harmonic                &                 85 &      75 &          20 &                  $\infty$ &      80 &          25 &                $\infty$ &    7965 &        1690 &             $\infty$ &   10000 &        3400 \\
Free + harmonic                &                 95 &      75 &          30 &                  $\infty$ &     110 &          30 &                $\infty$ &    1805 &        3895 &             $\infty$ &    1855 &        3900 \\
Free + harmonic                &                 25 &      20 &           5 &                 1285 &     460 &          10 &                $\infty$ &    5390 &        1060 &             $\infty$ &    7225 &        6385 \\
Free + harmonic                &                 80 &      95 &           5 &                  $\infty$ &     110 &          20 &                $\infty$ &    4380 &        3300 &             $\infty$ &    4800 &        4035 \\
Free + EM                      &                 90 &      85 &          20 &                  $\infty$ &    1190 &         115 &                $\infty$ &    6305 &        3380 &             $\infty$ &    6590 &        3435 \\
Free + EM                      &                125 &     120 &           0 &                 6240 &     885 &          70 &                $\infty$ &    7310 &        1865 &             $\infty$ &    7565 &        1865 \\
Free + EM                      &                115 &     115 &          15 &                15260 &     600 &          70 &                $\infty$ &    2430 &        1225 &             $\infty$ &    2845 &        4435 \\
Free + EM                      &                145 &      90 &           0 &                 6650 &     140 &           0 &                $\infty$ &    3000 &        5205 &             $\infty$ &    4530 &        8735 \\
Free + EM                      &                 80 &      80 &          10 &                  965 &     200 &          25 &                $\infty$ &    4635 &        1970 &             $\infty$ &    4690 &        2870 \\
Free + EM rational           &                 80 &      75 &           0 &                  $\infty$ &     580 &          70 &                $\infty$ &    5415 &        4150 &             $\infty$ &    5445 &        4175 \\
Free + EM rational           &                100 &     100 &          10 &                  $\infty$ &     460 &          45 &                $\infty$ &    2560 &         965 &             $\infty$ &    2575 &        5760 \\
Free + EM rational           &                140 &      95 &          10 &                11050 &     455 &          65 &                $\infty$ &    1960 &        1150 &             $\infty$ &    6295 &        4005 \\
Free + EM rational           &                120 &     100 &           5 &                13315 &     325 &         175 &                $\infty$ &    3970 &        1290 &             $\infty$ &    4335 &        3560 \\
Free + EM rational           &                 35 &      30 &          35 &                 1155 &     335 &          35 &                $\infty$ &    3245 &        2130 &             $\infty$ &    5115 &        5610 \\
Free + gravity + harmonic      &                150 &      75 &          25 &                 9085 &     130 &          30 &                $\infty$ &    3870 &        6145 &             $\infty$ &    5555 &        6185 \\
Free + gravity + harmonic      &                145 &      90 &           5 &                 6915 &     140 &          25 &                $\infty$ &    4525 &        3720 &             $\infty$ &   10275 &        4430 \\
Free + gravity + harmonic      &                105 &     100 &          15 &                 6925 &     155 &          40 &                $\infty$ &    6665 &        6560 &             $\infty$ &    8915 &        6845 \\
Free + gravity + harmonic      &                 95 &      95 &           5 &                  $\infty$ &     120 &          30 &                $\infty$ &    5790 &       10915 &             $\infty$ &   18450 &       13125 \\
Free + gravity + harmonic      &                135 &      95 &          15 &                 7970 &     190 &          45 &                $\infty$ &   13125 &        7045 &             $\infty$ &     $\infty$ &         $\infty$ \\
Free + gravity + EM            &                130 &     100 &          20 &                  $\infty$ &     575 &          40 &                $\infty$ &    3215 &        5095 &             $\infty$ &    3215 &        5100 \\
Free + gravity + EM            &                125 &     110 &          15 &                 5650 &     160 &          30 &                $\infty$ &    6085 &        4720 &             $\infty$ &    8025 &        4980 \\
Free + gravity + EM            &                 80 &      65 &          15 &                  $\infty$ &     630 &         120 &                $\infty$ &    4100 &        6250 &             $\infty$ &    4100 &        6570 \\
Free + gravity + EM            &                 80 &      75 &           5 &                  $\infty$ &      90 &          45 &                $\infty$ &    5910 &        5815 &             $\infty$ &    7295 &        6090 \\
Free + gravity + EM            &                 80 &      85 &          20 &                  $\infty$ &    1380 &         465 &                $\infty$ &    2390 &       11425 &             $\infty$ &    7450 &       11510 \\
Free + harmonic + EM           &                 85 &      75 &          25 &                  $\infty$ &     600 &         150 &                $\infty$ &    3775 &        4525 &             $\infty$ &    4675 &        5070 \\
Free + harmonic + EM           &                 85 &      90 &          25 &                  $\infty$ &    1245 &         200 &                $\infty$ &    6225 &        2340 &             $\infty$ &    6390 &        3180 \\
Free + harmonic + EM           &                115 &      85 &          15 &                16600 &     190 &          35 &                $\infty$ &    6035 &        1515 &             $\infty$ &   10065 &        2110 \\
Free + harmonic + EM           &                 80 &      70 &          35 &                  $\infty$ &     720 &         195 &                $\infty$ &    6990 &        3895 &             $\infty$ &    6995 &        6115 \\
Free + harmonic + EM           &                 85 &      65 &          10 &                  $\infty$ &     985 &         165 &                $\infty$ &    5660 &        1670 &             $\infty$ &    5820 &        1820 \\
Free + gravity + harmonic + EM &                 90 &      75 &           0 &                  $\infty$ &     540 &         255 &                $\infty$ &    8320 &        7390 &             $\infty$ &    9770 &        7590 \\
Free + gravity + harmonic + EM &                 95 &      80 &          15 &                  $\infty$ &    1265 &         635 &                $\infty$ &    6520 &        6365 &             $\infty$ &    8475 &        6475 \\
Free + gravity + harmonic + EM &                130 &      85 &          10 &                 8620 &     575 &         105 &                $\infty$ &    6320 &        4035 &             $\infty$ &    9705 &        7685 \\
Free + gravity + harmonic + EM &                 75 &      80 &           0 &                  $\infty$ &     815 &         425 &                $\infty$ &    7575 &        8405 &             $\infty$ &   10440 &        8620 \\
Free + gravity + harmonic + EM &                 80 &      65 &          20 &                  $\infty$ &     735 &         280 &                $\infty$ &    6715 &        4555 &             $\infty$ &   12495 &        8495 \\
\hline
Median                           &                 95 &      83 &          15 &                $\infty$ &     330 &          45 &                $\infty$ &    5403 &        3895 &            $\infty$&    6590 &        5100 \\
Mean                             &                 98 &      82 &          15 &                 $\infty$ &     455 &         109 &                $\infty$ &    5217 &        4171 &            $\infty$ &    6892 &        5499 \\
\hline
\end{tabular}
}
\end{center}
\caption{Same as previous table, but showing number of training epochs required to reach various MSE prediction accuracies. We record the metrics every 5 epochs, so all the epochs are multiples of 5. Note that the AI Physicist has superseded $10^{-2}$ MSE already by 0 epochs for some environments, showing that thanks to the lifelong learning strategy which proposes previously learned theories in novel environments,  reasonably good predictions can sometimes be achieved even without gradient descent training.
\label{DetailedResultsTable2}
}
\end{table*}

\subsection{Double Pendulum}
\label{appendix:double_pendulum}

Our double pendulum is implemented as two connected pendulums with massless rods of length 1 and that each have a point charge of 1 at their end. As illustrated in \fig{PendulumFig}, the system state is fully determined by the 4-tuple
$\y=(\theta_1, \dot{\theta}_1,\theta_2, \dot{\theta}_2)$ and immersed in a piecewise constant electric field $\E$:
$\E=(0, -E_1)$ in the upper half plane $y\geq-1.05$, and 
$\E=(0, E_2)$
in the lower half plane $y<-1.05$, using coordinates where $y$ increases vertically and the origin is at the pivot point of the upper rod.

We generate 7 environments by 
setting $(E_1, E_2)$ equal to $(E_0,2E_0)$, $(E_0,1.5E_0)$, $(E_0,E_0)$, $(E_0,0.5E_0)$, $(2E_0,E_0)$, $(1.5E_0,E_0)$, and $(0.5E_0,E_0)$, where $E_0=9.8$.
We see that there are two different EOMs for the double pendulum system depending on which of the two fields the lower charge is in (the upper charge is always in $E_1$).
We use Runge-Kutta numerical integration to simulate $\y=(\theta_1, \dot{\theta}_1,\theta_2, \dot{\theta}_2)$ for 10,000 time steps with interval of 0.05, and the algorithms' task is to predict the future $(\y_{t+1}$) based on the past 
($\x_t\equiv\y_t$; history length $T= 1$), and simultaneously discover the two domains and their different EOMs unsupervised. %Since each domain EOM is nonlinear and quite complicated, it is not possible to evaluate the symbolic matching. But we can still evaluate the unsupervised classification accuracy as in Appendix \ref{appendix:mystery_worlds}

In this experiment, we implement prediction function of the Baseline and Newborn both as a $N_{lay}^f$-layer neural net (we use $N_{lay}^f=6$) during DDAC. For the Newborn, each hidden layer has $N_{neur}^f=160$ neurons with hyperbolic tangent (tanh) activation, and for the Baseline, each hidden layer has $N_{neur}^f=320$ neurons with tanh activation for a fair comparison. For the Newborn, the optional AddTheories($\mathbfcal{T},D$) (step s8 in Alg. \ref{alg:divide_and_conquer}) is turned off to prevent unlimited adding of theories. The initial number $M$ of theories for Newborn is set to $M=2$ and $M=3$, each run with 10 instances with random initialization. Its domain classifier $\c=(c_1,c_2,...c_M)$ is a $N_{lay}^c$-layer neural net (we use $N_{lay}^c=3$) where each hidden layer has $N_{neur}^c=6$ neurons and leakyReLU activation. The last layer has linear activation.

\begin{table*}
\begin{center}
{\small
\begin{tabular}{| l|l | l| c c c |}
\hline
&Hyperparameter&Environments& Baseline & Newborn & AI Physicist \\
\hline
$\gamma$ & Generalized-mean-loss exponent & All & -1 & -1 & -1\\
$\beta_\f$& Initial learning rate for $\f_\theta$&All& 0.005 & 0.005 & 0.005 \\
$\beta_c$&Initial learning rate for $\c_\phi$ &All& 0.001 & 0.001 & 0.001 \\
$K$& Number of gradient iterations &All& 10000  & 10000 & 10000\\
$\sigma_c$ & Hidden layer activation function in $\c_\phi$ & All & - & leakyReLU & leakyReLU\\
$N_{\rm lay}^c$& Number of layers in $\c_\phi$ &All & - & 3 & 3 \\
$C$ &Initial number of clusters in theory unification  & All & 4 & 4 & 4\\
$\epsilon_{MSE}$ & MSE regularization strength  & All & $10^{-7}$  & $10^{-7}$ & $10^{-7}$ \\
\hline
$\epsilon_{L1}$ & Final $L_1$ regularization strength  & Mystery worlds & $10^{-8}$  & $10^{-8}$ & $10^{-8}$ \\
 &   & Double Pendulum & $10^{-7}$  & $10^{-7}$ & $10^{-7}$ \\
\hline
$N_{\rm lay}^f$ & Number of layers in $\f_\theta$ &Mystery worlds & 3 & 3 & 3 \\
 &  &Double Pendulum & 6 & 6 & 6 \\
\hline
$N_{\rm neur}^f$&
Number of neurons in $\f_\theta$ &Mystery worlds & 16 & 8 & 8 \\
&						&Double Pendulum & 320 & 160 & - \\
\hline
$N_{\rm neur}^c$&
Number of neurons in $\c_\phi$ &Mystery worlds & - & 8 & 8 \\
&						&Double Pendulum & - & 6 & - \\
\hline
$T$&
Maximum time horizon for input &Mystery worlds & 2 & 2 & 2 \\
&						&Double Pendulum & 1 & 1 & 1 \\
\hline
$\sigma_f$& Hidden layer activation function in $\f_\theta$ &Mystery worlds  & leakyReLU  & linear  & linear \\
&						&Double Pendulum  & tanh  & tanh  & - \\
\hline
$M_0$&
Initial number of theories&Mystery worlds &1 &2&2\\
&					&Double Pendulum&1 &2 \& 3&-\\
\hline
$M$& Maximum number of theories&Mystery worlds & 1 & 4 & 4\\
&					&Double Pendulum& 1 & 2 \& 3 & - \\
\hline
$\epsilon_{\rm add}$& MSE threshold for theory adding &Mystery worlds & - & $2 \times 10 ^ {-6}$ & $2 \times 10^{-6}$\\
&					&Double Pendulum& - & $\infty$ & - \\
\hline
$\eta_{\rm insp}$& Inspection threshold for theory adding &Mystery worlds & - & 30\% & 30\%\\
&					&Double Pendulum& - & $\infty$ & - \\
\hline
$\eta_{\rm split}$& Splitting threshold for theory adding &Mystery worlds & - & 5\% & 5\%\\
&					&Double Pendulum& - & $\infty$ & - \\
\hline
$\eta_{\rm del}$& Fraction threshold for theory deletion &Mystery worlds & - & 0.5\% & 0.5\%\\
&					&Double Pendulum& - & 100\% & - \\
\hline
\end{tabular}
}
\caption{
Hyperparameter settings in the numerical experiments. For a fair comparison between Baseline and the other agents that can have up to 4 theories, the number of neurons in each layer of Baseline is larger  so that the total number of parameters is roughly the same for all agents. The Baseline agent in Mystery worlds has leakyReLU activation to be able to able to account for different domains.
\label{hyperparameter_table}
}
\end{center}
\end{table*}

\clearpage
\bibliography{aiphysicist}

\end{document}